\documentclass[sigconf]{acmart}

\copyrightyear{2018} 
\acmYear{2018} 
\setcopyright{acmcopyright}
\acmConference[CCS '18]{2018 ACM SIGSAC Conference on Computer and Communications Security}{October 15--19, 2018}{Toronto, ON, Canada}
\acmBooktitle{2018 ACM SIGSAC Conference on Computer and Communications Security (CCS '18), October 15--19, 2018, Toronto, ON, Canada}
\acmPrice{15.00}
\acmDOI{10.1145/3243734.3243737}
\acmISBN{978-1-4503-5693-0/18/10}

\usepackage{booktabs} 

\usepackage[ruled]{algorithm2e} 

\SetAlFnt{\small}
\SetAlCapFnt{\small}
\SetAlCapNameFnt{\small}
\SetAlCapHSkip{0pt}
\IncMargin{-\parindent}

\usepackage{amsmath} 
\usepackage{amssymb} 
\usepackage{amsfonts} 
\usepackage{mathtools} 
\usepackage{url}
\usepackage[colorinlistoftodos]{todonotes}
\usepackage{fancyhdr}
\usepackage{graphicx}
\usepackage{epstopdf}
\usepackage{array}
\usepackage{caption}
\usepackage{subcaption}
\usepackage{makecell}

\usepackage{cleveref}
\newcommand{\imagesPath}{}

\newcommand{\negspace}{\vspace{0\baselineskip}}
\newcommand{\snegspace}{\vspace{0\baselineskip}}

\newcounter{todocounter}

\def\btc{%
  \leavevmode
  \vtop{\offinterlineskip 
    \setbox0=\hbox{B}%
    \setbox2=\hbox to\wd0{\hfil\hskip-.03em
    \vrule height .3ex width .15ex\hskip .08em
    \vrule height .3ex width .15ex\hfil}
    \vbox{\copy2\box0}\box2}}

\newcommand{\currentTime}{t}
\newcommand{\totalNumberOfRigs}{k}
\newcommand{\totalNumberOfPlayers}{n}

\newcommand{\rigIndex}{j}
\newcommand{\otherRigIndex}{l}
\newcommand{\playerIndex}{i}
\newcommand{\otherPlayerIndex}{j}

\newcommand{\difficultyPerRig}{\mu \left( \rigStartTimeVector \right)}
\newcommand{\timeBetweenBlocks}{\ensuremath{ \textit{Block\_Interval} }\xspace}
\newcommand{\expectedTotalFees}{\ensuremath{ \textit{Expected\_Total\_Fees} }\xspace}

\newcommand{\baseReward}{\lambda_0}
\newcommand{\capexPerRigPerTime}{\ensuremath{ c_\textit{cap} }\xspace}
\newcommand{\opexPerRigPerTime}{\ensuremath{ c_\textit{op} }\xspace}
\newcommand{\feeIncPerTime}{\lambda_{t}}
\newcommand{\baseRewardRatio}{\ensuremath{ \textit{EBRR} }\xspace}

\newcommand{\rigsOfPlayerName}{R}
\newcommand{\rigStartTimeName}{s}

\newcommand{\rigsOfPlayerWithIndex}[1]{\rigsOfPlayerName{}_{#1}}

\newcommand{\rigStartTimeVectorWithIndex}[1]{\rigStartTime{}{#1}}

\newcommand{\optimizer}{equilibrium-search-tool}

\newcommand{\allRigsIndices}{  \left\{ 1,2,\ldots,\totalNumberOfRigs \right\}}
\newcommand{\listOfRigIndices}{\left\{ \rigIndex{}_1 ,  \ldots , \rigIndex{}_m \right\}}

\newcommand{\rigStartTime}[2]{\rigStartTimeName{}_{#2}^{#1}}
\newcommand{\rigStartTimeVector}{\bar{\rigStartTimeName{}}}

\newcommand{\sumForItemInSet}[3]{\sum_{#1 \in #2}^{} #3 }
\newcommand{\sumofTimeSinceItemsInSet}{\sumForItemInSet
{\rigIndex}
{\allOperatingRigsAllPlayers{}}
{(\currentTime-\rigStartTime{}{\rigIndex})}}
\newcommand{\sumofTimeSinceItemsInSetAtTimeWithCurrentTime}[1]{\sumForItemInSet
{\rigIndex}
{\allOperatingRigsAllPlayersAtSomeStartTimeWithRigIndex{#1}}
{(\currentTime-\rigStartTime{}{\rigIndex})}}
\newcommand{\sumofTimeSinceItemsInSetAtTimeWithVal}[2]{\sumForItemInSet
{\rigIndex}
{\allOperatingRigsAllPlayersAtSomeStartTimeWithRigIndex{#1}}
{(#2-\rigStartTime{}{\rigIndex})}}

\newcommand{\rigRelativeStartTime}[2]{\widetilde{\rigStartTimeName{}}_{#2}^{ #1}}
\newcommand{\rigRelativeStartTimeFrac}[2]{{#1} / {#2}}
\newcommand{\rigRelativeStartTimeEqFrac}[4]{\ifthenelse{\equal{#3}{0}}{\rigRelativeStartTime{#1}{#2}=0}{\rigRelativeStartTime{#1}{#2}=\rigRelativeStartTimeFrac{#3}{#4}}}

\newcommand{\blockFindingRandomEvent}{\ensuremath{ \textit{B}}\xspace}
\newcommand{\blockFindingRandomEventByRigIndex}{\blockFindingRandomEvent{}_\rigIndex}

\newcommand{\relativePowerOfPlayer}{\alpha_{\playerIndex{}}(\currentTime)}

\newcommand{\sizeOfSet}[1]{\left\vert{#1}\right\vert}

\newcommand{\allOperatingRigsFunctionName}{\ensuremath{ \textit{active}}\xspace}

\newcommand{\normProfitFunctionName}{\ensuremath{\textit{utility}}\xspace}

\newcommand{\allPlayersIndex}{}

\newcommand{\incomeFunctionName}{\ensuremath{ \textit{Income}_{\playerIndex}}\xspace}
\newcommand{\costFunctionName}{\ensuremath{ \textit{Expenses}_{\playerIndex}}\xspace}
\newcommand{\profitFunctionName}{\ensuremath{ \textit{Profit}_{\playerIndex}}\xspace}

\newcommand{\costPlayerI}{\expectedValueOfRandVarDependent{\costFunctionName}{\blockFindingRandomEvent{}=\currentTime{}}}
\newcommand{\incomePlayerI}{\expectedValueOfRandVarDependent{\incomeFunctionName}{\blockFindingRandomEvent{}=\currentTime{}}}
\newcommand{\profitPlayerI}{\expectedValueOfRandVarDependent{\profitFunctionName}{\blockFindingRandomEvent{}=\currentTime{}}}

\newcommand{\probability}[1]{\text{Pr(}#1\text{)}}
\newcommand{\expectedProfitPlayerI}{\expectedValueOfRandVar{\profitFunctionName{}}}
\newcommand{\normExpectedProfitPlayerI}{\ensuremath{\textit{\normProfitFunctionName}}_{\playerIndex}\xspace}

\newcommand{\expectedValueOfRandVar}[1]{\ensuremath{\textit{E}}\xspace \left({#1} \right)}
\newcommand{\expectedValueOfRandVarDependent}[2]{\expectedValueOfRandVar{{#1} \mid {#2}}}

\newcommand{\startTimesIntervalSetName}{R}
\newcommand{\startTimesIntervalSet}{
\startTimesIntervalSetName = 
\left\{
\left( \rigStartTime{}{1},\rigStartTime{}{2} \right),
\left( \rigStartTime{}{2},\rigStartTime{}{3} \right),
\ldots,
\left( \rigStartTime{}{\totalNumberOfRigs{}-1},\rigStartTime{}{\totalNumberOfRigs{}} \right),
\left( \rigStartTime{}{\totalNumberOfRigs{}},\infty \right) 
\right\}  
}

\newcommand{\allOperatingRigsPlayerI}{\ensuremath{ \allOperatingRigsFunctionName{}_{\playerIndex{}}\text{(}\currentTime{}\text{)}}\xspace}
\newcommand{\allOperatingRigsAllPlayers}{\allOperatingRigsFunctionName{}_{\allPlayersIndex{}}\text{(}\currentTime{}\text{)}}

\newcommand{\allOperatingRigsAllPlayersAtSomeStartTimeWithRigIndex}[1]{\allOperatingRigsFunctionName{}_{\allPlayersIndex{}}
\left (#1 \right )}

\newcommand{\allOperatingRigsAtSomeStartTime}[1]{\allOperatingRigsFunctionName{}
\left ({#1} \right)}

\newcommand{\sumOfTimeDifferencesPlayerI}{\sumForItemInSet{\rigStartTime{}{}}
{\allOperatingRigsPlayerI{}}
{(\currentTime-\rigStartTime{}{})}}

\newcommand{\probabilityTimeLessThanBlockIndex}{\probability{\currentTime \le
\blockFindingRandomEventByRigIndex}}
\newcommand{\probabilityTimeGreaterThanBlockIndex}{\probability{\currentTime \ge
\blockFindingRandomEventByRigIndex}}

\newcommand{\probabilityTimeLessThanBlock}{\probability{\currentTime \le
\blockFindingRandomEvent}}

\newcommand{\pdfFunctionName}{f}
\newcommand{\pdfRigIndex}{\pdfFunctionName_{\blockFindingRandomEventByRigIndex}(
\currentTime; \rigStartTime{}{\rigIndex} , \difficultyPerRig)}
\newcommand{\pdfRigMin}{\pdfFunctionName_{\blockFindingRandomEvent}(\currentTime;
\rigStartTimeVector , \difficultyPerRig)}

\newcommand{\pdfRigMinAtInf}{\lim_{\currentTime\to\infty}
\pdfFunctionName_{\blockFindingRandomEvent}(\currentTime; \rigStartTimeVector , \difficultyPerRig)}

\newcommand{\cdfFunctionName}{F}
\newcommand{\cdfRigIndex}{\cdfFunctionName_{\blockFindingRandomEventByRigIndex}(
\currentTime; \rigStartTime{}{\rigIndex} , \difficultyPerRig)}
\newcommand{\cdfRigMin}{\cdfFunctionName_{\blockFindingRandomEvent}(\currentTime;
\rigStartTimeVector , \difficultyPerRig)}

\newcommand{\cdfRigMinAtZero}{\cdfFunctionName_{\blockFindingRandomEvent}(0;
\rigStartTimeVector , \difficultyPerRig)}
\newcommand{\cdfRigMinAtInf}{\lim_{\currentTime\to\infty}
\cdfFunctionName_{\blockFindingRandomEvent}(\currentTime; \rigStartTimeVector , \difficultyPerRig)}

\newcommand{\minOfBlockFindingRandomEventByRigIndex}{\blockFindingRandomEvent
= \min_{  \rigIndex \in \allRigsIndices }
\blockFindingRandomEventByRigIndex}

\newcommand{\probabilityTimeLessThanAllBlockIndices}{
\bigcap_{\rigIndex \in \allRigsIndices } 
\probability{\currentTime \le \blockFindingRandomEventByRigIndex}
}
\newcommand{\prodOfProbabilityTimeLessThanAllBlockIndices}{
\prod_{\rigIndex=1}^{\totalNumberOfRigs} \probability{\currentTime
\le \blockFindingRandomEventByRigIndex}}

\newcommand{\meanValOfBlockFindingRandomEvent}{ E [ \blockFindingRandomEvent ]}

\newcommand{\SETTINGHIGHOP}{\ensuremath{ \textit{high\_op}}\xspace}
\newcommand{\SETTINGMEDOP}{\ensuremath{ \textit{med\_op}}\xspace}
\newcommand{\SETTINGLOWOP}{\ensuremath{ \textit{low\_op}}\xspace}

\newcommand{\PROBSETTINGALLZERO}{\ensuremath{ \textit{classical}}\xspace}
\newcommand{\PROBSETTINGALLHALF}{\ensuremath{ \textit{uniform gap}}\xspace}
\newcommand{\PROBSETTINGASCATTER}{\ensuremath{ \textit{arbitrary gap}}\xspace}

\newcommand{\numberOfSimIterationsEachSetting}{10}
 
\newcommand{\FIGSCALE}{0.45}

\newcommand{\VERTICALSUBFIGSCALE}{0.4}

\begin{document}

\title{The Gap Game}


\author{Itay Tsabary} 
\affiliation{Technion}
\email{sitay@campus.technion.ac.il}

\author{Ittay Eyal} 
\affiliation{Technion}
\email{ittay@technion.ac.il}

\begin{abstract} 
Blockchain-based cryptocurrencies secure a decentralized consensus protocol by incentives. 
The protocol participants, called miners, generate (mine) a series of blocks, each containing monetary transactions created by system users. 
As incentive for participation, miners receive newly minted currency and transaction fees paid by transaction creators.
Blockchain bandwidth limits lead users to pay increasing fees in order to prioritize their transactions. 
However, most prior work focused on models where fees are negligible. 
In a notable exception, Carlsten et al.~\cite{arvindcutoff} postulated that if incentives come only from fees then a mining gap would form~--- miners would avoid mining when the available fees are insufficient. 

In this work, we analyze cryptocurrency security in realistic settings, taking into account all elements of expenses and rewards. 
To study when gaps form, we analyze the system as a game we call \emph{the gap game}. 
We analyze the game with a combination of symbolic and numeric analysis tools in a wide range of scenarios. 

Our analysis confirms Carlsten et al.'s postulate; indeed, we show that gaps form well before fees are the only incentive, and analyze the implications on security. 
Perhaps surprisingly, we show that different miners choose different gap sizes to optimize their utility, even when their operating costs are identical. 
Alarmingly, we see that the system incentivizes large miner coalitions, reducing system decentralization. 
We describe the required conditions to avoid the incentive misalignment, providing guidelines for future cryptocurrency design. 
\end{abstract}



\maketitle

\noindent
\framebox{
	\parbox{\dimexpr\linewidth-2\fboxsep-2\fboxrule}{\itshape%
		This paper is now in proceedings of the 2018 ACM SIGSAC conference on computer and communications security (CCS '18).
		
		For citations, please refer to the official CCS '18 paper page at \href{https://dl.acm.org/citation.cfm?id=3243737}{https://dl.acm.org/citation.cfm?id=3243737}.
	}
}

    \section{Introduction}

Since their introduction in 2008~\cite{nakamoto2008bitcoin}, blockchain protocols are securing rapidly increasing amounts of money in the form of so-called cryptocurrencies. 
As of today, the market cap of the first cryptocurrency, Bitcoin, is estimated at \$160B~\cite{blockchain2018marketCap,coinmarketcap}, and the market cap of all cryptocurrencies, most of which are secured with blockchain protocols, is estimated at \$350B~\cite{blockchain2018marketCap,coinmarketcap}. 

Cryptocurrencies facilitate users transactions of a currency internal to the system.
The underlying protocol, the blockchain, is operated by independent principals called~\emph{miners}. 
Miners collect transactions in blocks and append them to the blockchain, forming a globally-agreed order of transactions. 
Instead of relying on a central control, the most prominent blockchain-based cryptocurrencies~\cite{nakamoto2008bitcoin,bch2018site,buterin2013ethereum,litecoin2013site,sasson2014zerocash} rely on utilizing incentives to secure the system. 
They use~\emph{proof of work} (\emph{PoW})~\cite{Dwork1993,jakobsson1999proofs,nakamoto2008bitcoin}, requiring participants to solve moderately-difficult cryptographic puzzles to generate blocks. 
The idea is that to successfully attack the system, one would need to control resources proportional in amount to those of all participating miners. 
To motivate miner participation, cryptocurrencies incentivize them with \emph{block rewards} comprising \emph{subsidy}, newly minted currency created at the generation of each new block, and \emph{transaction fees}, paid by the transactions. 
Preliminaries on cryptocurrencies and blockchain protocols are in Section~\ref{sec:background}. 


In the dominant operational cryptocurrency systems~\cite{nakamoto2008bitcoin,buterin2013ethereum}, the subsidy is the substantial part of the incentive as of today. 
And indeed, despite the breadth of research on blockchain security~\cite{mukhopadhyay2016brief, bonneau2015sok, den2017economists, tama2017critical}, and despite the significance of incentives for blockchain security, most prior work studied the incentives scheme when the reward comes only from subsidy~\cite{analysisOfBlockChainProtocol,schrijvers2016incentive,eyal2014majority,eyal2016ng,sapirshtein2016optimal,eyal2015miner,kwon2017selfish}. 
However, as a cryptocurrency gains traction, the incoming load of transactions increases~\cite{blockchain2018mempoolCount,ether2018pendingTx}.
Since transaction bandwidth is limited, a fee market forms~-- users offer higher fees to motivate miners to place their transaction quickly~\cite{blockchain2018txFees,etherscan2018txFees}.
Moreover, in Bitcoin and several other cryptocurrencies minting rate decays over time. 
Hence, fees are on the path to become a substantial part of cryptocurrency rewards. 

Carlsten et al.~\cite{arvindcutoff} postulated that in a certain scenario, a \emph{mining gap} would form. 
Their model assumes only operational expenses and no subsidy, and that block size is unbounded, so miners place all pending transactions when mining a block. 
Therefore, once a block is generated, there are no unclaimed transactions and therefore no unclaimed fees, and so no incentive to mine the next block until sufficient fees have accrued, resulting in a gap in mining period. 
In Section~\ref{sec:relatedWork} we review previous work.


In this work, we analyze the incentives and equilibrium of blockchain-based PoW cryptocurrency systems, taking into account rewards from both subsidy and fees, and both capital and operational expenses. 
We present our model in Section~\ref{sec:Model}.

This model gives rise to a game we call~\emph{the gap game} (Section~\ref{sec:theGapGame}). 
It is played among the miners, which compete on finding blocks~-- the first to find a block gets rewarded, while all suffer expenses.
It is a one-shot game, where the miners decide when to start their mining rigs, and strive to optimize their average revenue, maximizing the difference between their income and expenses. 

To study the game properties we first develop some tools (Section~\ref{sec:analysisTools}). 
We develop expressions for the average time to find a block and the average revenue of a miner given the start times of all players. 
Our results match scenarios analyzed in prior art for subsidy-only rewards~\cite{nakamoto2008bitcoin,analysisOfBlockChainProtocol} and fee-only rewards~\cite{arvindcutoff}, and an independent simulation for scenarios not previously analyzed, where miners choose different mining gaps. 
We then proceed to derive the utility function of each player, and a numerical analysis tool to find equilibria in the game.
Since the expressions for miner utility do not lend themselves to symbolic analysis, we use numerical analysis to find~$\varepsilon$-Nash Equilibria over a wide range of parameters. 

Our analysis reveals several things (Section~\ref{sec:analysisResults}). 
As predicted~\cite{arvindcutoff}, a mining gap does form when subsidy is sufficiently small and operational expenses are large. 
Unexpectedly, we show that mining gaps varies between miners based on size, even if their per-rig properties are identical. 
Additionally, we show that by forming coalitions miners increase their gains. 
The implication is that in a system where fees are sufficiently large miners are incentivized to form coalitions, leading centralization and defeating the basic premise of a blockchain system. 
We therefore find the required system parameters to ensure the avoidance of a mining gap and its detrimental effects. 
These values can be used to inform the design of incentive mechanisms in current and future cryptocurrency systems.
We conclude by estimating when Bitcoin will be prone to these effects.

We conclude in Section~\ref{sec:conclusion} with a discussion on the implications of our results on operational cryptocurrencies and on future cryptocurrency design. 
Note these results apply for active cryptocurrency systems, such as Bitcoin, Zcash, Litecoin and so forth, while also very much relevant for the design of new systems.

In summary, we make the following contributions: 
\begin{itemize} 
\item Derive expressions for miner revenue with gaps, 
\item Define the gap game, played among miners, 
\item Analyze equilibria in a variety of settings,  
\item Find that gaps differ among miners, 
\item Find that miners profit by forming coalitions, 
\item Estimate when Bitcoin will be affected, and 
\item Show how to prevent those predicaments.
\end{itemize}

\section{Background}
\label{sec:background}
\snegspace

Blockchain-based cryptocurrency systems~\cite{nakamoto2008bitcoin,buterin2013ethereum,litecoin2013site,sasson2014zerocash} allow users to exchange currency via in-system transactions, without the verification of a centralized authority.
Such systems use a public distributed ledger, named the blockchain, to record all internal transactions performed.
When a user creates a transaction, it is propagated across the cryptocurrency network, and eventually all other users are familiar with it.
The ledger is composed of blocks, a set of transactions grouped together.
Participants who run the blockchain protocol, named miners, add new blocks to the blockchain.

The aforementioned cryptocurrency systems operate in a~\emph{permissionless} setting, allowing any participant to join or leave the network.
A challenge of operating in this setting is to keep security and fairness, as malicious participants can join the network and might deviate from honest behavior.
Hence, as part of their protocol, systems use different methods to ensure the desired honest behavior of their participants.
A popular method is proof of work~\cite{Dwork1993,jakobsson1999proofs} that requires a miner to solve a moderately-difficulty cryptographic puzzle in order to create a valid block.
By solving such puzzle, a miner proves she invested computational work.
Systems that use proof of work rely on the assumption that at least $50\%$ of computational work invested on mining is by honest participants~\cite{nakamoto2008bitcoin}.
If malicious users control more than $50\%$ of the computational power, they can employ double-spending attacks~\cite{karame2012two}.
The system is designed to assure that participants are incentivized to follow the protocol rules, and failing to do so will result in decreased profit.

The mining process for a new block goes as follows.
A miner groups a set of transactions to be included in the new block and validates them using the blockchain.
Then, she looks for a solution for a cryptographic puzzle, which is based on the of selected transactions, the last block added to the blockchain and the cryptocurrency protocol.
Attached a valid solution, the block is propagated in the network and other miners agree to add it to their blockchains.
When a miner adds a new block to her blockchain, she restarts the mining process with respect to that new block.

In a permissionless setting, computational power may join and leave the system.
Therefore, the block time interval might vary, which is undesired.
To avoid this predicament, the system's protocol defines a fixed block time interval, and adjusts a difficulty parameter, which determines the difficulty of the cryptographic puzzles.
If blocks are created at higher (lower) rate than desired, the protocol sets the difficulty to increase (decrease) the time required for solving the cryptographic puzzle.

Miners attempt to create a valid block by iteratively guessing solutions for the cryptographic puzzle.
The process of guessing a solution can be modeled as a Bernoulli trial~--- a solution is guessed, randomly resulting in a 'success' if the solution fits (and then a valid block is created), or by a 'failure' if it doesn't.
The success rate of each trial is fixed and determined by the aforementioned difficulty parameter. 
Observing a series of such trials, the required number of trials for a success result is geometrically distributed.
Therefore, the time required for a successful result is drawn from the exponential distribution.

Note that both the geometric and the exponential distributions are memoryless, so the number of previously failed trials or the time that already passed do not change future probability of success. 
As a result, miner's chance of finding a valid solution is not changed by how many solutions it had attempted previously\footnote{This holds for any practical matter as the solutions space is practically infinite.}.
Hence, if a miner re-picks the set of transactions to be included in the block, and by doing so restarts the mining process, her chances of mining the next block are not decreased.

Mining blocks comes with a cost.
Mining rigs, the machines used for the mining process, require electricity for their operation~\cite{bitcoinEnergyIndex,etherEnergyIndex}.
Hardware maintenance, network connection and real-estate, all are required to operate rigs and all carry expenses for miners~\cite{miningRigsPrices,realWorldBitcoinMine,bitcoinMiningCostsCountryComparison}.
To incentivize participants to mine, systems offer rewards in the form of currency.
The rewarded currency comes from two sources~--- newly minted currency that's created as a part of a valid block, and transaction fees paid by transactions included in the block.
The amount of minted currency is determined by the cryptocurrency protocol, and the amount of fees is determined by the set of transactions the miner included in the block.
Each allocated transaction may offer a different fee, and miners get to pick which transactions they want to include in their blocks.

In two of the most popular cryptocurrency system nowadays, Ethereum and Bitcoin, the reward is dominant by the minted currency. 
In Ethereum, roughly $20k$ new Ethers are minted daily~\cite{etherscan2018etherSupply} while fees pay about $2k$ daily~\cite{etherscan2018txFees}.
In Bitcoin, the expected daily subsidy is $\btc 12.5 \cdot 24\cdot6= \btc 1800$, as average of $6$ blocks are generated every hour, each minting $\btc 12.5$.
The daily paid fees varies and averages around a few hundreds BTCs a day~\cite{blockchain2018txFees}.

This trend will eventually change, as allocated transaction fees are on the rise.
Blocks are bounded, and miners have to pick the set of transactions to include.
Many transactions end up not being picked at all.
Users who wish to get their transactions picked by miners increase the paid fees to incentivize picking their transactions.
Another cause for the expected trend change is that many cryptocurrencies, including Bitcoin, are designed to mint a finite supply of currency.
The monetary idea of the finite supply is to prevent inflation.
In Bitcoin for example, approximately every four years, the amount of of newly minted coins from new blocks is halved.
The expected number of total Bitcoins is estimated to be roughly $21$ millions~\cite{wiki2018controlledSupply,bitcoinsAfterAllMined,bitcoinsLimitedAmount}.

\section{Related work}
\label{sec:relatedWork}
\snegspace

Most of the previous work on cryptocurrencies incentives focused on models where the block reward is composed mostly of subsidy.
In the original Bitcoin white paper~\cite{nakamoto2008bitcoin}, fees are mentioned briefly and an intuitive reasoning about incentives is presented. 
Kroll et al.~\cite{kroll2013economics} analyze Bitcoin as a consensus game when fees are sufficiently low and conclude their impact is negligible.
Eyal and Sirer~\cite{eyal2014majority} show a deviant mining strategy named~\emph{selfish mining}, by which an attacker increase her relative reward.
Sapirshtein et al.~\cite{sapirshtein2016optimal} and Nayak et al.\cite{nayak2015stubborn} both show more sophisticated variations of the original selfish mining attack that increase the attacker's reward when applied.
Other work by Eyal~\cite{eyal2015miner} shows mining pools are incentivized to allocate some of their mining rigs to infiltrate other mining pools.
Once an infiltrating rig finds a block for the attacked pool, it withholds rather than publishing it.
The work shows that an equilibrium exist where two pools infiltrate one another, in which they both end up losing compared to if they were not attacking to begin with.
Kwon et al.~\cite{kwon2017selfish} combine the infiltration attack with selfish mining.
In their work, the infiltrating rig selectively alternates between performing withholding and selfish mining attacks.
All these works consider a model where the subsidy is the dominant incentive for mining and expenses are negligible.
In this work we use a different model, where miners have expenses that differ according to their mining strategy.
We also consider the profit of a miner is comprised of both subsidy and fees.

Babaioff et al.~\cite{babaioff2012baloons} discuss incentives for propagating transactions in a cryptocurrency network.
They offer and analyze several reward schemes to incentivize participants to distribute transactions in the network.
In this work we analyze systems with the traditional reward scheme, where participants are rewarded for mining blocks.
Transaction propagation is not incentivized, as in the classical reward scheme.

M{\"o}ser and B{\"o}hme~\cite{bitcoinTransactionFeesReview} review and analyze the history of transaction fees in Bitcoin. 
They conclude that historically miners prefer to follow the protocol rules rather than optimize their gains.
They predict such state is sustainable only when fees are a negligible part of the incentive. 
In our work, we analyze systems where fees are not negligible and show how such systems incentivize participants to undesired behavior.

Carlsten et al.~\cite{arvindcutoff} analyze Bitcoin when the mining incentive comes solely from fees, in a model where the number of transactions that can be placed in a block is unbounded.
In their model there is no residual fee after block generation as all transactions are included in the previous block, and so the block reward immediately after a block is found is zero. 
They analyze mining strategies and show how miners are incentivized to fork the main chain, disturbing security and liveness.
They also revise selfish mining and show an improved version suited dominant fees incentive.
An interesting conjecture briefly presented in their work is of the formation of a mining gap, a period of time in which miners turn their mining rigs off to reduce mining expenses.
When such mining gap exists, the mining power utilization of the network is suboptimal.
In proof of work scheme the immediate implication is that the system is less resilient to attacks.
In this work, we present a model to analyze miners' profits and use it to show that mining gaps do form.
Our model holds for both bounded and unbounded blocks, as well as for combinations of subsidy and fees as part of the block reward.
In their work, the mining gap conjecture was for a set of identical miners that all stop and start mining simultaneously.
In contrast, we show that different miners prefer different mining gaps.
We also show that rational miners are ought to form coalitions to increase their gains, leading to a centralized system.
We analyze loss of resilience to attacks.
We conclude by showing that with sufficient initial block reward, all miners are incentivized to resort to the default mining strategy.

Biais et al.~\cite{biais2018blockchain} analyze the investment in mining equipment required by miners in proof of work cryptocurrencies.
They show that miners require excessive acquisition of mining equipment to stay competitive with other miners. 
In this work we assume the mining equipment acquired is fixed for the network, yet we consider it as part miners' expenses.

Fruitchain~\cite{pass2017fruitchains} is a protocol that is $\epsilon$-Nash incentive compatible against any minority coalition. 
It shows that if fees are evenly distributed across different blocks as fees are smeared, the potential increase from deviating from the protocol is bounded. 
Hybrid Consensus~\cite{pass2016hybrid}, Sleepy consensus~\cite{pass2017sleepy} and Solida~\cite{abraham2016solida} are all newer protocols for implementing distributed consensus with blockchains.
They presume an altruistic majority of participants and do not consider incentives.
Algorand~\cite{gilad2017algorand} is another such protocol that use~\emph{proof of stake} instead of proof of work.
It explicitly does not consider incentives, which call for a different definition in the proof of stake scheme.
Ouroboros Praos~\cite{kiayias2017ouroboros} is also a proof of stake blockchain protocol.
It uses a new reward mechanism aimed to mitigate block withholding attacks. 
Bitcoin-NG~\cite{eyal2016bitcoin} is a new protocol with the intentions of scaling Bitcoin.
It utilizes proof of work for picking a leader, who creates microblocks to validate transactions.
Rewards are distributed by consecutive leaders, yet it also assumes both negligible fees and miner expenses.
Lavi et al.~\cite{lavi2017redesigning} considers two new bidding schemes for Bitcoin's fees market, while focusing on incentivizing miners to offer their true bids rather than strategically bid.
Our results focus on Bitcoin-like cryptocurrency protocols and with Bitcoin's current incentive scheme, and do not trivially apply to these other protocols.

\section{Model}
\label{sec:Model}
\snegspace

We present a realistic model of cryptocurrency systems that we use throughout the rest of this work. 
As commonly done in blockchains analysis~\cite{eyal2014majority,arvindcutoff,pass2017fruitchains, garay2015backbone}, we model systems in a quasi-static state.
That means no miners join or leave~\cite{huberman2017monopoly,kroll2013economics}, existing miners maintain their behavior and the system reached equilibrium.  
Therefore, in our model the system comprises a fixed set of miners and a fixed set of mining rigs.
Each miner controls at least one rig and each rig is controlled by exactly one miner.
We assume for simplicity that mining rigs are identical~\cite{arvindcutoff}.
Rigs have two states~--- off, the default state, and on.
Each miner assigns a \emph{start time} for each of her controlled rigs, in which the rig is turned on.
We often refer to a turned-on rig as an~\emph{active} rig.

Once a rig is turned on, the time it takes to find a valid block is exponentially distributed with a fixed rate parameter, which is shared among all rigs~\cite{nakamoto2008bitcoin,sapirshtein2016optimal,nayak2015stubborn,eyal2014majority}.
Therefore the time to find the first block by any of the rigs is the minimum of all finding times by all different rigs.
The value of the rate parameter is determined by the cryptocurrency protocol such that the expected block time interval is of a constant value that is also determined by the protocol.
The rate parameter represents the difficulty of the cryptographic puzzle, and we use the terms difficulty and rate interchangeably.
The assigned start times of rigs by miners affect the value of the rate parameter.
If blocks are found too fast (too slow), then the difficulty parameter value is changed by the protocol to decrease (increase) the rate of each individual rig.
In equilibrium, the rate parameter is of a fixed value.

The rig that finds the block first awards its controlling miner the block reward, which is comprised of two parts.
The first part is \emph{fees reward} that comes from aggregation of newly introduced transactions to the system.
This reward is time-dependent, as the time progresses there are more pending transactions in the system, and the potential fees reward grows. 
The second part is a subsidy we refer to as \emph{base reward}, which to the contrary of the fees reward is fixed over time.
This reward is comprised of the minting of new currency with the creation of each block, as well as the expected reward from transaction fees considering the expected initial set of pending transactions.
Note that the finding of a new block does not reward any other miners except the miner who found it.

To participate in the system miners expend resources, and we differentiate two types of such resources.
First, miners have capital expenses (capex), which are for owning a rig~\cite{miningRigsPrices,realWorldBitcoinMine} and apply whether the rig is active or not.
Miners also have operational expenses (opex), which are paid for having a rig actively mining~\cite{etherEnergyIndex,bitcoinEnergyIndex} i.e. owning an active rig.
Note that these expenses apply for all miners and not just on those who manage to successfully mine blocks.

Once a block is found, all miners move on to find the next block.
This process is repeated indefinitely.
The profit of a miner for each block is the difference between her total expenses and her total reward.
Rational miners strive to maximize their profits, giving rise to a game.

\section{The Gap Game}
\label{sec:theGapGame}
\snegspace

The repeated search for the blocks becomes a series of independent one-shot competitions, in each only one miner gets the reward but all miners pay expenses.
To reason about expected revenues, rather than considering the individual iterations we consider a one-shot game played by the miners. 
A player's strategy is the choice of start times of all of her rigs~-- when each rig is turned on.
The choice of start times are made a-priori by all players.
We define the utility of a player to be her expected profit, which is her expected income minus her expected expenses.

The system comprises~$\totalNumberOfRigs$ mining rigs controlled by~$\totalNumberOfPlayers$ players.
Player~$\playerIndex$ controls the set of rigs with indices~$\rigsOfPlayerWithIndex{\playerIndex}$. 
Note that 
$ \forall \rigsOfPlayerWithIndex{\playerIndex} :
\rigsOfPlayerWithIndex{\playerIndex} \ne \emptyset $,
$ \forall 
\playerIndex \ne \otherPlayerIndex :
\rigsOfPlayerWithIndex{\playerIndex} \cap
\rigsOfPlayerWithIndex{\otherPlayerIndex}  = \emptyset $ 
and
 $  \bigcup_{\playerIndex=1}^{\totalNumberOfPlayers}
 {\rigsOfPlayerWithIndex{\playerIndex}} = \allRigsIndices $.
Denote the expected block time interval achieved by the protocol by~$\timeBetweenBlocks$.
The start time of each rig~$\rigIndex$ is~$\rigStartTime{}{\rigIndex}$, and we denote the normalized start time~$\rigRelativeStartTime{}{\rigIndex} = \tfrac{\rigStartTime{}{\rigIndex}}{\timeBetweenBlocks}$. 
Once a rig is turned on, the time it requires to find a block is exponentially distributed with a rate parameter $\difficultyPerRig$.

For simplified writing in the following section, we denote $\rigStartTimeVector$ as the vector of increasing order $\totalNumberOfRigs{}$ rigs' start times.

All rigs are identical~-- Each mining rig costs~\capexPerRigPerTime per time unit for ownership and~\opexPerRigPerTime per time unit if it is turned on. 

The utility of player $\playerIndex$, denoted~$\normExpectedProfitPlayerI$, is its expected profit, namely its expected rewards minus its expected expenses. 
We derive an expression for the utility function in the following section.

The utility is affected by the strategies of all players.
As is common in miner behavior analysis~\cite{kroll2013economics,eyal2014majority,sapirshtein2016optimal,eyal2015dilemma,kwon2017selfish}, our solution concept is myopic~--- each player chooses her best-response strategy based on the current strategies of other players.
Players do not take into consideration how other players will adapt based on their new choice of strategy.

\paragraph{Strategy space}
Note that the strategy space does not include turning rigs off, as this is an irrational behavior.
Block finding time of an active rig is drawn from the exponential distribution, which is memoryless.
That means the probability for a rig to find the block in some time interval is not affected by how much time had already passed since that rig began mining.
Therefore, a single rig's chances of finding a block are not decreasing over time.
Recall that the total reward also increase over time.
Hence, if at some point in time the reward justified turning a rig on, then this justification holds from that time until the block is found.

\snegspace
\subsection{Parameters Analysis}
\label{sec:analysisParameters}
\snegspace

The parameters values are affected by a wide range of factors, stemming from different sources.
The fees are affected by the system users and the market~\cite{blockchain2018txFees,blockchain2018marketCap,etherFeesGuide,predictBitcoinFees,howToCalcTranFee,bitcoinTransactionFeesReview}.
The base reward is also affected by systems user and market, as these affect the residual fees, but also by the minting rate, which is defined by the cryptocurrency protocol.
Capex is affected by factors such as technological advancements of mining rigs efficiency~\cite{biais2018blockchain}, personnel wages, and real estate costs~\cite{cheapestPlacesMiningBitcoin,bitcoinMiningFarm,realWorldBitcoinMine}.
Opex is affected primarily by the electricity costs~\cite{bitcoinMiningCostsCountryComparison,cheapestPlacesMiningBitcoin,realWorldBitcoinMine,etherEnergyIndex,bitcoinEnergyIndex} for operating the mining rigs.
That includes both the actual puzzle solving process as well as cooling expenses.
These parameters are therefore not only difficult to estimate, but they vary between different currencies, and also over time for the same currency. 
Hence, we analyze the system for a range of parameters values to make general observations, focusing on trends that are robust across the parameter range.

We begin by analyzing how fees accumulate in the system, and then move towards determining parameter values which we'll be used throughout the rest of this work.

\subsubsection{Fees Reward Accumulation over Time}

Accurately predicting the fees accumulation function of the pending transactions is out of the scope of this work, and we resort to an educated approximation. 
We measure how fees accumulate over time in the Bitcoin network and apply our findings to the general model.

We conducted the following measurement at February~2018.
Using a Bitcoin node connected to the Bitcoin network, we monitor the pending transactions awaiting to be included in blocks.
At fixed time intervals of one second, we find the most rewarding set of transactions to include in a valid~1~MB Bitcoin block.
We record the fees that transactions in this set offer.
The values recorded correspond to blocks~509426 up to~509605 and span about~30 hours of measurements.

\begin{figure}[!t]
   \centering
  \includegraphics[width=\FIGSCALE\textwidth]{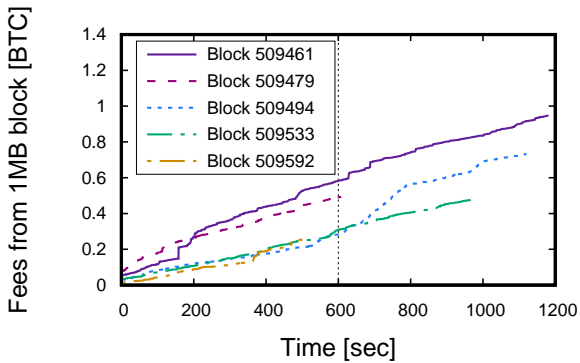} 
  \caption{Fees in most rewarding 1MB block accumulation in Bitcoin's mempool.}
 \label{fig:feesAccumulationInBitcoin}
\end{figure}

In Figure~\ref{fig:feesAccumulationInBitcoin} we present the potential fees reward as a function of time, during the time it took to mine a specific block, for some arbitrary measured blocks.
The vertical dashed line shows the expected block time interval, which is~600 seconds in Bitcoin.
As expected, some blocks required more (less) time than the expected interval.

Using linear regression on all the measured blocks, we calculate the squared correlation value and get an average of $R^2=0.96$.
We conclude a linear approximation is reasonable and therefore treat the fees reward as if it increases linearly.

We also note that immediately after a block is found, there are still pending transactions awaiting to be included in future blocks.
We can consider the expected fees of these pending transactions as if they were part of the fixed base reward out of the total block reward.
 
Hence, we model the total block reward as a linear function, where the slope is the expected fees accumulation rate, and the intercept is the sum of the newly minted currency and the expected fees available immediately after a block is found.
We repeated these measurements at other dates for different periods of time and received similar results.
We denote $\feeIncPerTime$ as the fees accumulation rate and $\baseReward$ as the base reward.

\subsubsection{Analysis Parameters}

We denote by $\expectedTotalFees$ the expected total fees accumulating during the expected time to find a block, namely, $\expectedTotalFees = \timeBetweenBlocks \cdot \feeIncPerTime$.
Denote by $\baseRewardRatio$ the ratio of the expected base reward and the expected accumulated fees, so $\baseRewardRatio= \tfrac{\baseReward}{\expectedTotalFees}$. 
Throughout the following sections we present results for different values of $\baseRewardRatio$.

For all experiments we choose the following parameters arbitrarily: 
Fees increase rate is set to $\feeIncPerTime = 1$, the expected block interval to $\timeBetweenBlocks = 10000$, and the number of rigs to $\totalNumberOfRigs = 128$.

\begin{table}[!t]
\centering
\begin{tabular}{ |c|c|c| } 
 \hline
\textbf{Name} &  \textbf{$\opexPerRigPerTime $}  & \textbf{$\capexPerRigPerTime $} \\
 \hline
$\SETTINGHIGHOP$ & 0.02 & 0.00 \\ 
 \hline 
$\SETTINGMEDOP$ &  0.01 & 0.01 \\ 
 \hline
$\SETTINGLOWOP$ &  0.00 & 0.02 \\ 
 \hline
\end{tabular}
\caption{Opex and capex settings.} 
\label{tab:opexCapexSettings} 
\end{table}

Recall we analyze systems at a quasi-static state and miners do not join or leave the system.
The profit for miners should therefore be slightly more than the interest rate plus associated risk.
For simplicity, to avoid introducing unnecessary parameters, we set the expense parameters such that the expected profit of miners will be zero.
Therefore, we choose values so $\opexPerRigPerTime + \capexPerRigPerTime$ is of a fixed value.

The ratio between of the two types of expenses, opex and capex, can vary considerably among cryptocurrencies. 
Different cryptocurrencies use different proof of work~\cite{alwen2018sustained,miller2014permacoin,zhang2017rem,gervais2016security,intel2018sawtooth} with different computational costs, varying mining technology~\cite{miningRigsPrices,biais2018blockchain,realWorldBitcoinMine}, and varying electricity expenses~\cite{bitcoinMiningCostsCountryComparison,cheapestPlacesMiningBitcoin}.
Therefore, we use three different settings that are of interest for the ratio of capex and opex parameters values, which are detailed below and summarized in Table~\ref{tab:opexCapexSettings}. 
Two settings describe extreme cases, where in one all the expenses are opex, and in the other all the expenses are capex.
The third setting describes the average case of the first two, where the opex and capex are equal.

The system's properties are determined by the parameters ratios~--- the ratio of expected fees reward and the base reward, the ratio of opex and opex and so forth.
Throughout the rest of this work we cover a wide range of these ratios that demonstrate the important trends. 
We emphasize that different values satisfying the same ratios yielded the same qualitative results.

\section{Game Analysis}
\label{sec:analysisTools}
\snegspace

To find the utility of each player, we start by analyzing the block finding time probability distribution. 
This is a function of the players' selection of start times.  
We model the block finding time as a random variable denoted~$\blockFindingRandomEvent$ with cumulative distribution function (CDF) and probability density function (PDF) denoted $\cdfRigMin$ and $\pdfRigMin$, respectively.

\begin{figure}
		\centering
        \begin{subfigure}[b]{\VERTICALSUBFIGSCALE\textwidth}
                \includegraphics[width=\linewidth]{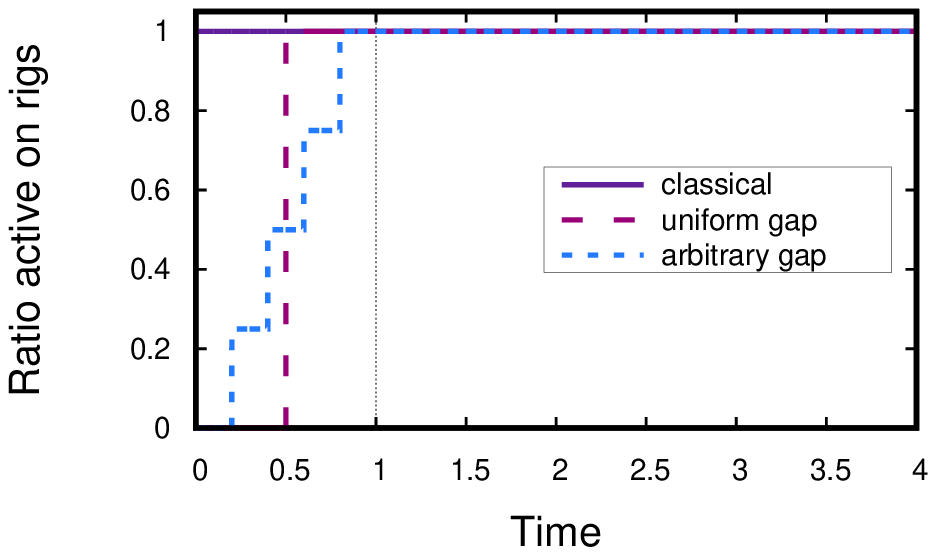}
                \caption{ Ratio of turned on rigs.}
                \label{subfig:probabilitiesActive}
        \end{subfigure}%

        \begin{subfigure}[b]{\VERTICALSUBFIGSCALE\textwidth}
                \includegraphics[width=\linewidth]{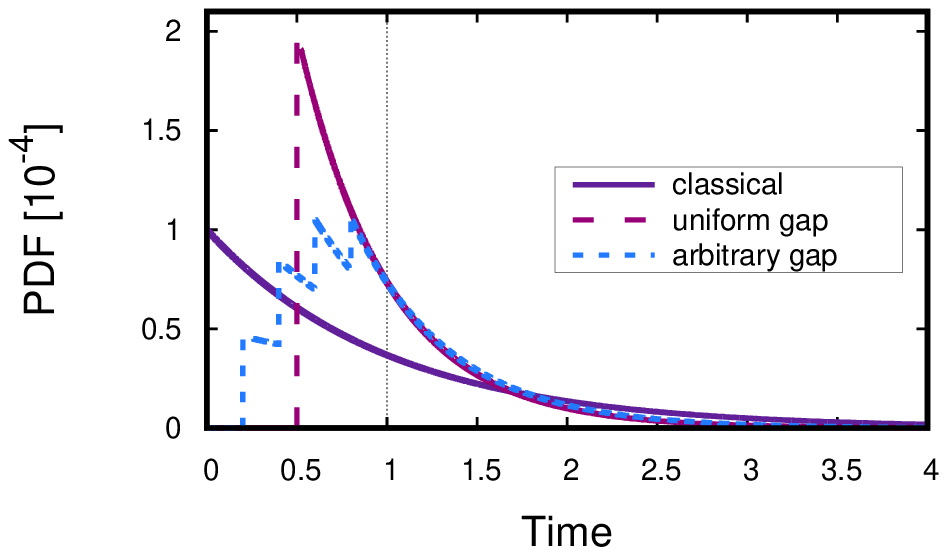}
                \caption{PDF~---$ \pdfRigMin $}
                \label{subfig:probabilitiesPDF}
        \end{subfigure}%

        \begin{subfigure}[b]{\VERTICALSUBFIGSCALE\textwidth}
                \includegraphics[width=\linewidth]{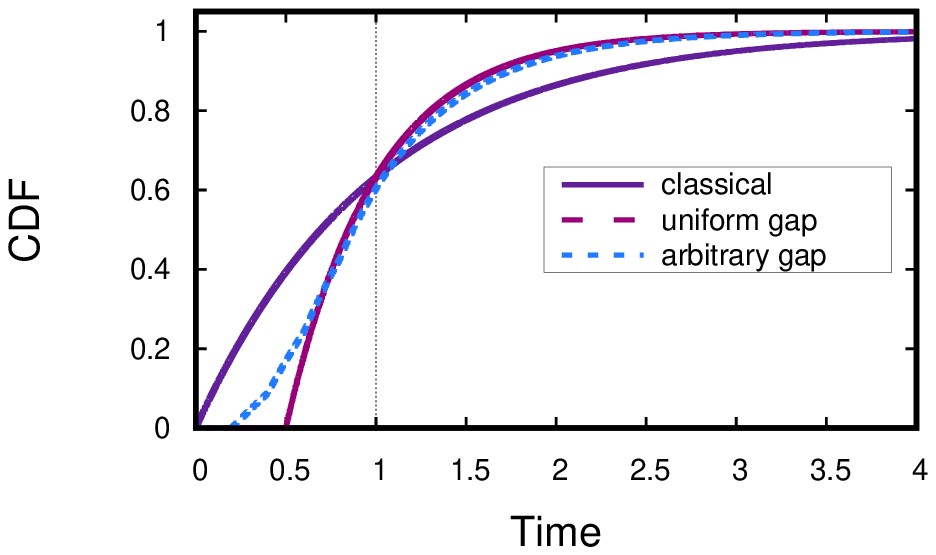}
                \caption{CDF~---$  \cdfRigMin $}
                \label{subfig:probabilitiesCDF}
        \end{subfigure}%
        \caption{System properties for different scenarios.}
        \label{fig:probabilities}
\end{figure}

We begin by discussing the difference of the probability distributions in our model distributions from ones considered in prior art. 
We present three different scenarios of rigs' start time choices and the derived probabilities of the system.
Table~\ref{tab:probabilitiesSetting} lists the values used in each scenario and Figure~\ref{fig:probabilities} depicts the resultant distributions. 
Figure~\ref{subfig:probabilitiesActive} shows the ratio of active rigs as a function of time, while Figures~\ref{subfig:probabilitiesPDF},~\ref{subfig:probabilitiesCDF} show the PDF and CDF of the block finding time $\blockFindingRandomEvent$, respectively.
In this example the expected block interval is set to be $\timeBetweenBlocks=1$.
Each scenario has four equal-size players, each controlling~32 out of the total $\totalNumberOfRigs = 128$ rigs in the system.

In the $\PROBSETTINGALLZERO$ scenario, all players set their rigs' start times to~0. 
This is the scenario commonly analyzed in the literature~\cite{nakamoto2008bitcoin,sapirshtein2016optimal,nayak2015stubborn,eyal2014majority}. 
In Figure~\ref{subfig:probabilitiesActive} we see a constant ratio of~1 as all rigs are set to have $\currentTime=0$.
From Figures~\ref{subfig:probabilitiesPDF},~\ref{subfig:probabilitiesCDF} we learn that  $\pdfRigMin > 0, \cdfRigMin>0$ for all  $\currentTime$, which is expected as all rigs are active throughout the entire scenario. 
As all block finding times of single rigs are exponentially distributed, the block finding time is also exponentially distributed.
The rate parameter is $\tfrac{1}{\timeBetweenBlocks}$ such that the expected time will be $\timeBetweenBlocks$.

In the $\PROBSETTINGALLHALF$ scenario, all players set their rigs' start times to~0.5. 
This is scenario is analyzed in~\cite{arvindcutoff}. 
In Figure~\ref{subfig:probabilitiesActive} we can see that all the ratio is 0 while $\currentTime < 0.5$ and to 1 while $\currentTime \ge 0.5$ , as all rigs are set to have $\rigStartTime{}{}=0.5$.
In Figures~\ref{subfig:probabilitiesPDF},~\ref{subfig:probabilitiesCDF} $\pdfRigMin = 0, \cdfRigMin = 0$ while $\currentTime < 0.5$ as no rigs are active.  
When $\currentTime \ge 0.5$ , all rigs are turned on and $\pdfRigMin > 0, \cdfRigMin > 0$.
In this case, the block finding times of single rigs are shifted-exponentially distributed, the block finding time is also shifted-exponentially distributed.
The shift is of~0.5 time units and the rate parameter is doubled $\tfrac{2}{\timeBetweenBlocks}$ to compensate.
Notice that the expected block time interval is still $\timeBetweenBlocks$.

In the $\PROBSETTINGASCATTER$ scenario, each player set her rigs' start times to a different value.
To the best of our knowledge, this scenario is first analyzed in this work.
In Figure~\ref{subfig:probabilitiesActive} we can the ratio increases as time progresses.
The spikes occur at the times where rigs are turned on.
Notice that for $\currentTime < 0.2$ all the rigs are still turned off and the ratio is~0.
At $\currentTime = 0.2,0.4,0.6,0.8$ , the change in the number of turned on rigs causes the CDF in Figure~\ref{subfig:probabilitiesCDF} to be semi-differentiable, resulting in the jump discontinuities of the PDF in Figure~\ref{subfig:probabilitiesPDF}.
As expected, $\pdfRigMin = 0, \cdfRigMin = 0$ while $\currentTime < 0.2$ as no rigs are turned on. 
Note that for all scenarios $ \pdfRigMinAtInf = 0$, $\cdfRigMinAtZero = 0$ and $ \cdfRigMinAtInf = 1$.

\begin{table}
\centering
\begin{tabular}{ |c|c|c|c|c| } 
\hline
				  &  \multicolumn{4}{c|}{\textbf{Rigs Quarter}} \\
\hline
\textbf{Scenario}			  &  \textbf{Q1} & \textbf{Q2} & \textbf{Q3} & \textbf{Q4} \\
\hline
$\PROBSETTINGALLZERO$~\cite{nakamoto2008bitcoin,sapirshtein2016optimal,nayak2015stubborn,eyal2014majority} & 0 &  0  & 0 & 0 \\
\hline
$\PROBSETTINGALLHALF$~\cite{arvindcutoff} & 0.5 & 0.5 & 0.5 & 0.5  \\
\hline
$\PROBSETTINGASCATTER$ & 0.2 & 0.4 & 0.6 & 0.8 \\
\hline
\end{tabular}
\captionof{table}{Rigs Start Times.} 
\label{tab:probabilitiesSetting} 
\end{table}

The rest of this section is organized as follows.
In Section~\ref{sec:derivingProbabilityFunctions} we derive an expression for the distribution based on the selected values of $\rigStartTimeVector{}$, and proceed to derive an expression for $\normExpectedProfitPlayerI$ in Section~\ref{sec:utilanalysis}.
Then, in Section~\ref{sec:simIntro} we present a simulator designed to confirm our theoretical analysis.
We conclude in Section~\ref{sec:systemEquilibrium} by presenting an optimizing tool created to find equilibria in the game.

\snegspace
\subsection{Distribution Analysis}
\label{sec:derivingProbabilityFunctions}
\snegspace
The first step towards analyzing the system is to derive an expression for the distribution, namely $\cdfRigMin$ and $\pdfRigMin$, based on players' strategies.
We begin by deriving the distribution of a single rig.
Observe any single rig $\rigIndex$ that with start time $\rigStartTime{}{\rigIndex}$.
Denote the time this rig requires for finding a block as a random variable $\blockFindingRandomEventByRigIndex$.
Recall that the rate of a single rig is $\difficultyPerRig$, which is set by the protocol. 
The value of $\blockFindingRandomEventByRigIndex$ is drawn from the shifted exponential distribution, with a shift of $\rigStartTime{}{\rigIndex}$ and rate $\difficultyPerRig$.

The PDF of $\blockFindingRandomEventByRigIndex$ is 
\begin{equation*}
   \pdfRigIndex =
\begin{cases}
    0, 								  & \currentTime \le \rigStartTime{}{\rigIndex} \\
    \difficultyPerRig \cdot   \exp\left({-\difficultyPerRig
    (\currentTime-\rigStartTime{}{\rigIndex}) }\right)   & \currentTime >
    \rigStartTime{}{\rigIndex}
\end{cases}
\end{equation*}
and its CDF is
\begin{equation*}
   \cdfRigIndex = 
\begin{cases}
    0, 								  & \currentTime \le \rigStartTime{}{\rigIndex} \\
    1 -   \exp\left({-\difficultyPerRig 
    (\currentTime-\rigStartTime{}{\rigIndex}) }\right)   & \currentTime >
    \rigStartTime{}{\rigIndex} \end{cases}\,.
\end{equation*}

As $\cdfRigIndex = \probabilityTimeGreaterThanBlockIndex = 1 -
\probabilityTimeLessThanBlockIndex$ 
we get that
\begin{equation*}
  \probabilityTimeLessThanBlockIndex = 
\begin{cases}
    1, 								  & \currentTime \le \rigStartTime{}{\rigIndex} \\
      \exp\left({-\difficultyPerRig 
    (\currentTime-\rigStartTime{}{\rigIndex}) }\right)   & \currentTime >
    \rigStartTime{}{\rigIndex} \end{cases}\,.
\end{equation*}

All rigs are competing on finding the next block.
The rig that finds the next block first is the rig with the minimal value of $\blockFindingRandomEventByRigIndex$.
Therefore, the time required for finding the next block is $\minOfBlockFindingRandomEventByRigIndex$.

We define for any time $\currentTime$ and any player $\playerIndex$ the set $\allOperatingRigsPlayerI$ to be all player $\playerIndex$'s rig indices that are active at time $\currentTime$ :
$\allOperatingRigsPlayerI = \left\{
\rigIndex \mid
\rigIndex \in \rigsOfPlayerWithIndex{\playerIndex} 
\land
\rigStartTime{}{\rigIndex} \le \currentTime
\right\}$.
We define $\allOperatingRigsAllPlayers$ to be the set of all active rigs at time $\currentTime$.
Note that
$\allOperatingRigsAllPlayers =
  \bigcup_{\playerIndex=1}^{\totalNumberOfPlayers} {\allOperatingRigsPlayerI}
$.

The probability that none of the rigs have found a block by time~$\currentTime$, $\probabilityTimeLessThanBlock$, is the product of $\probabilityTimeLessThanBlockIndex$ for all $\rigIndex$ (as rigs are independent of another one).
This probability is given by
\begin{multline*}
  \probabilityTimeLessThanBlock = 
  \probabilityTimeLessThanAllBlockIndices =\\
   \prodOfProbabilityTimeLessThanAllBlockIndices =
	\exp\left({-\difficultyPerRig \cdot \sumofTimeSinceItemsInSet{}  }\right)\,. 
\end{multline*}

The CDF of $\blockFindingRandomEvent$ is therefore
\begin{equation}
\label{eq:cdfBlock}
  \cdfRigMin = 1 - \probabilityTimeLessThanBlock =
1- \exp\left({-\difficultyPerRig \sumofTimeSinceItemsInSet{} }\right)
\end{equation}

and the derivative is its PDF,
\begin{equation}
\label{eq:pdfBlock}
  \pdfRigMin = 
    \difficultyPerRig \cdot \sizeOfSet{\allOperatingRigsAllPlayers{}} \cdot
    \exp\left({-\difficultyPerRig \cdot \sumofTimeSinceItemsInSet{}}\right)\,.
\end{equation}

As expected, when $\sizeOfSet{\allOperatingRigsAllPlayers{}} = 0$ then $\sumofTimeSinceItemsInSet{} = 0$ which results in $ \cdfRigMin = 0$ and $\pdfRigMin = 0$.
We can verify that $\pdfRigMin$ is a valid PDF by checking that $\int_{-\infty}^{\infty} \pdfRigMin d \currentTime =1$ holds.
We prove this is in fact the case in Appendix~\ref{app:validPdf}.
We also find the value of $\difficultyPerRig$ at equilibrium.
This process is presented in Appendix~\ref{app:calcDiff} and utilized when required throughout this work.

\snegspace
\subsection{Utility}
\label{sec:utilanalysis}
\snegspace
We are now ready to express $\normExpectedProfitPlayerI$.
Recall that $\normExpectedProfitPlayerI$ is the \emph{expected} profit of player $\playerIndex$.
We define three new random variables~--- $\incomeFunctionName$ , $\costFunctionName$ , $\profitFunctionName$, representing the income, expenses and profit of player $\playerIndex$, respectively.
Throughout the rest of this section, we assume the value of $\blockFindingRandomEvent$ is $\currentTime$, and use it to find the expected profit of player $\playerIndex$ that is denoted as $\profitPlayerI$. 
We then use the law of total expectation and the PDF of $\blockFindingRandomEvent$ from Equation~\ref{eq:pdfBlock} to derive an expression for $\expectedProfitPlayerI$, which is by definition $\normExpectedProfitPlayerI$.  

\snegspace
\subsubsection{Income Function}
\snegspace
We model the income function linearly with a slope of $\feeIncPerTime$ and an intercept of $\baseReward$.
Therefore, the total available reward at time~$\currentTime$ is $\baseReward + \feeIncPerTime \cdot \currentTime$.

Recall that once a rig is turned on, the time it requires to find a block is drawn from the exponential distribution.
The exponential distribution is memoryless, meaning the time that passed does not affect the chances of a rig to find the block.
Since the rate parameter $\difficultyPerRig$ is shared among all rigs, at any given time all the active rigs have the same chance to find the block, regardless of how much time they had been active for already.

Observe the set of active rigs at the time the block is found $\allOperatingRigsAllPlayers$.
The probability of a specific active rig to find the block is one divided by the total number of active rigs.
Note that since the block was found at time~$\currentTime$, then $\exists \rigIndex \in \allRigsIndices $ such that $ \rigStartTime{}{\rigIndex}\le\currentTime$ and therefore $\sizeOfSet{\allOperatingRigsAllPlayers} > 0$.
Players control many rigs, so the probability that player $\playerIndex$ controls the rig that found the block is the number of her controlled active rigs divided by the total number of active rigs.
We denote the ratio of player $\playerIndex$'s active rigs out of all the active rigs at time~$\currentTime$ as $\relativePowerOfPlayer =\tfrac{\sizeOfSet{\allOperatingRigsPlayerI}}{\sizeOfSet{\allOperatingRigsAllPlayers}}$.
The ratio $\relativePowerOfPlayer$ is therefore the expected factor of player $\playerIndex$'s portion of the total reward.

We conclude that if a block was found at time~$\currentTime$, then the expected income of player $\playerIndex$ is
\begin{equation}
\label{eq:income}
   \incomePlayerI = 
   \relativePowerOfPlayer (\baseReward +
   \feeIncPerTime \cdot \currentTime)\,.
\end{equation}
\snegspace
\subsubsection{Expenses Function}
\snegspace
Recall that players have two kind of expenses. 
The first, capex, for owning a rig.
The second, opex, for having a rig active.

Capex applies for all rigs controlled by the player, whether they are turned on or not.
For each rig, the capex it imposes by time~$\currentTime$ is the product of $\capexPerRigPerTime$ and $\currentTime$.
Recall that $\rigsOfPlayerWithIndex{\playerIndex}$ is the set of rig indices that player $\playerIndex$ controls, which totals with $\sizeOfSet{\rigsOfPlayerWithIndex{\playerIndex}}$ rigs.
The total capex of player $\playerIndex$ by time $\currentTime$ are therefore $\capexPerRigPerTime \cdot \sizeOfSet{\rigsOfPlayerWithIndex{\playerIndex}} \cdot \currentTime$.

Opex applies only for active rigs.
For each active rig, the expenses it imposes by time~$\currentTime$ is the product of $\opexPerRigPerTime$ and the time duration this rig is turned on already.
At time~$\currentTime$, active rig $\rigIndex$ with $\rigStartTime{}{\rigIndex}$ has been active for $\currentTime - \rigStartTime{}{\rigIndex}$ time.
Summing for all rigs of player $\playerIndex$ results that by time $\currentTime$ the total opex are $\opexPerRigPerTime \cdot \sumOfTimeDifferencesPlayerI$.

Combining both of these expenses, if a block was found at time~$\currentTime$ then the expected expenses of player $\playerIndex$ are
\begin{equation}
\label{eq:cost}
   \costPlayerI = \capexPerRigPerTime \cdot
   \sizeOfSet{\rigsOfPlayerWithIndex{\playerIndex}}  \cdot
   \currentTime + \opexPerRigPerTime \cdot \sumOfTimeDifferencesPlayerI \,.
\end{equation}

\snegspace
\subsubsection{Profit Function}
\snegspace
The expected profit of a player is her expected income minus her expected expenses.
Using Equations~\ref{eq:income} and~\ref{eq:cost}, we get that if a block was found at time~$\currentTime$ then the expected profit of player $\playerIndex$ is
\begin{multline}
\label{eq:profit}
   \profitPlayerI =\\
    \incomePlayerI - \costPlayerI\,.
\end{multline}

\snegspace
\subsubsection{Utility Function}
\snegspace
To get the expected profit of a player, we use the law of total expectation (sometimes referred to as the \emph{smoothing theorem}).
We use the PDF of $\blockFindingRandomEvent$ that from Equation~\ref{eq:pdfBlock}.
Therefore, the expected profit of player $\playerIndex$, which is also defined as her utility, is
\begin{multline}
\label{eq:utility}
   \normExpectedProfitPlayerI =
   \expectedProfitPlayerI=
   \expectedValueOfRandVar{\profitPlayerI} = \\
   \int_{-\infty}^{\infty} \left(\profitPlayerI \cdot
   \pdfRigMin\right) d \currentTime\,.
\end{multline}

\snegspace
\subsection{Cryptocurrency System Simulator}
\label{sec:simIntro}
\snegspace
In addition to the theoretical analysis, we implemented a cryptocurrency system simulator. 
It is built as an event driven simulation and operates at the continuous time space.
It includes a set of miners that control mining rigs.
Each miner keeps a private copy of the blockchain and compete with the other miners on finding the next block.
We use exponentially distributed random events to simulate block mining intervals.
The rate parameter of the exponential distribution is set such that the mean block time interval is kept at a fixed value.
When a miner finds a block, he announces it to the rest of the miners.
Each miner sets a-priori a start time for each of her controlled rigs, which refers to required time to pass since the finding of the previous block so this specific rig will become active.
Active rigs keep on mining until the next block is found by any rig.
Transactions accumulate over time and found blocks include the allocated fees as a reward, as well as a base reward for each block.
Miners also pay expenses as a function of their controlled rigs (capex) and the time those rigs were turned on (opex).

We emphasize that the theoretical analysis yields the expected profit for a player from a \emph{single} block, while simulations create a long blockchain containing many blocks mined by all participating miners.
Hence, when referring to the results of the simulator, we refer to mean profit of a miner over time.
\snegspace
\subsection{System Equilibrium Search}
\label{sec:systemEquilibrium}
\snegspace

The utility presented in Equation~\ref{eq:utility} is derived given all players' strategies.
If a player changes her strategy, then the utility of all the other players is also affected.
We are interested in finding equilibria, i.e. strategies of all players such that no player can improve her utility by changing her strategy.

The utility of a player is infeasible to express in a symbolic manner.
It is a function of all player strategies as well as the difficulty parameter, which can be expressed only as an implicit function (in any case where there are at least two distinguished start times).
Therefore, we use numerical analysis to find equilibria in the system.

We implemented an \optimizer{}~--- a tool we use to numerically search for an equilibrium, and that works in the following manner.
The \optimizer{} receives as an input the system income and expenses parameters, as well as a list of tuples representing all players' strategies.
Each tuple of that list is in the form of 
$ \left\{ 
\playerIndex{}, 
\rigIndex{}, 
\rigStartTime{}{} 
\right \}$,
where $\playerIndex{}$ is a player's index, $\rigIndex{}$ is a rig index that are controlled by player $\playerIndex{}$ and $\rigStartTime{}{}$ is a start time selected for rig $\rigIndex{}$.
Note that $\listOfRigIndices{} \subset \allRigsIndices$.

Iteratively, the \optimizer{} chooses at random an input tuple
$ \left\{ 
\playerIndex{}, 
\rigIndex{}, 
\rigStartTime{}{} 
\right \}$,
and searches what value of a new start time for rig $\rigIndex{}$ will result in maximal utility for player $\playerIndex{}$. 
This process is repeated until no player increases her utility by changing any of her rigs, meaning an equilibrium is reached.

Note that all equilibria found by such process are only $\epsilon$-Nash-equilibria, as they are limited by the numerical precision of the calculation.
To counter that predicament, we repeat the search process with different random start times and different optimizing order.
In all conducted experiments, the randomness introduced had no effect on the output equilibrium.
That strengthens our analysis of an equilibria.

\section{Analysis Results}
\label{sec:analysisResults}
\snegspace

We study the system behavior in a wide range of scenarios, detailed in Section~\ref{sec:analysisParameters}~-- from the common case in today's operational currencies where subsidy dominates rewards to the extreme case where fees dominate rewards, and with varying expenses distributions. 
We proceed to verify our analysis tools using the cryptocurrency system simulator, compare it to known results, and observe some predictable trends (Section~\ref{sec:miningGapSim}). 

In Section~\ref{sec:caseStudies} we present the first trend that was not predicted in prior art~-- even when rig parameters are identical, players of different sizes choose different gap sizes in equilibrium.
We also present the utility of a single player as a function of other players' strategies and show players are expected to optimize. 
Then, in Section~\ref{sec:equalPlayersEq}, we analyze the game with equal-size players of varying size, showing how the gap game encourages equilibria that affect the security of the system.

We want to compare the utility of players in systems with different reward schemes.
To eliminate the effect of players having high utility as they are in systems that offer high rewards, we instead consider on the utility of players out of the total utility available in the system.
We also want to eliminate the effects of bigger players having more utility and therefore we actually consider the utility normalized by size.
More formally, we use \emph{normalized} utility of players, that is defined to be the utility presented in Equation~\ref{eq:utility}, normalized by two factors. 
The first factor is $\feeIncPerTime \cdot \timeBetweenBlocks + \baseReward$ that represents the total expected income from a block in the system.
In our experiments this factor varies as a function of~$\baseReward$.
This normalization allows us to compare systems with different $\baseReward$.
The second normalization factor is the number of rigs each player controls, which varies for each player.

\subsection{Analysis Tools Validation}
\label{sec:miningGapSim}

We validate our analysis by comparing our theoretical results with both simulated and previously known results.
In Section~\ref{sec:exp1RelProfit} we compare with the classical scenario discussed in previous work~\cite{nakamoto2008bitcoin,sapirshtein2016optimal,nayak2015stubborn,eyal2014majority}, when there are no gaps. 
Next, in Section~\ref{sec:exp2ActualProfit}, we present and analyze a scenario with arbitrary gaps.
For this scenario there is no previous work to compare with, so we compare our theoretical results only against the simulation.
\snegspace
\subsubsection{Scenario One~-- No Mining Gap}
\label{sec:exp1RelProfit}
\snegspace

We analyze a simple scenario where the system is comprised of two miners, both mine without a mining gap, and with no expenses.
We use the analytic expression and the simulator to obtain the \emph{relative} utility of player~1~-- the ratio of a her utility out of the utility of all players.
We vary the player~1's relative mining power and plot its relative utility. 
This is the common metric that was used in previous work~\cite{nakamoto2008bitcoin,sapirshtein2016optimal,nayak2015stubborn,eyal2014majority}.
In those previous works, reward from fees is negligible, meaning the relative utility is the ratio of blocks mined by a player.
It is also the metric used for the scenario where there is no reward from minting, all transactions are identical in their fees, and blocks are unbounded~--- the relative utility is the ratio of transactions included by blocks mined by the player~\cite{arvindcutoff}.
In both cases the expected result is for a player with~$\alpha$ relative mining power to have a relative utility of~$\alpha$.
All of these works neglect the expenses of players. 
Hence, for comparison purposes we nullify these expenses in this particular scenario by setting $\opexPerRigPerTime=\capexPerRigPerTime=0$.

We compare the relative utility according to the game analysis, the simulated results, and the expected result.
For the simulated results, we use the average of $\numberOfSimIterationsEachSetting$ different runs with different random seeds.
We use several values for~$\baseRewardRatio$ and, as expected, the results match. 

            \subsubsection{Scenario Two~-- Arbitrary Mining Gap}
            \label{sec:exp2ActualProfit}

\begin{table}[!t]
\centering
\begin{tabular}{ |c|c| } 
\hline
\textbf{\thead{Portion of \\ player 1's rigs}} & \textbf{\thead{Normalized \\ start time}} \\
\hline
 0.2 & 0.1 \\
\hline 
 0.7 & 0.3 \\
\hline
 0.1 & 0.9 \\ 
\hline
\end{tabular}
\quad
\begin{tabular}{ |c|c| } 
\hline
\textbf{\thead{Portion of \\ player 2's rigs}} & \textbf{\thead{Normalized \\ start time}} \\
\hline
 0.2 & 0.2 \\
\hline 
 0.4 & 0.5 \\
\hline
 0.4 & 0.6 \\ 
\hline
\end{tabular}
\captionof{table}{Normalized start times for players' rigs.} 
\label{tab:splitting} 
\end{table}

\begin{figure}
		\centering
        \begin{subfigure}[b]{\VERTICALSUBFIGSCALE\textwidth}
                \includegraphics[width=\linewidth]{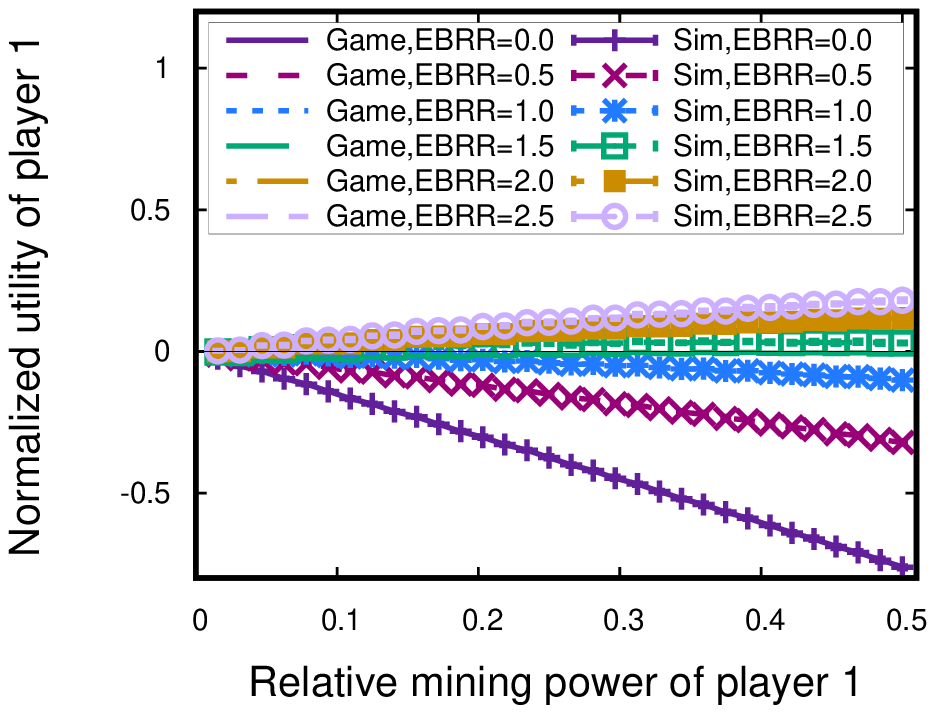}
                \caption{$ \SETTINGLOWOP $}
                \label{subfig:SimulationTheoryProfitComparisonA}
        \end{subfigure}%

        \begin{subfigure}[b]{\VERTICALSUBFIGSCALE\textwidth}
                \includegraphics[width=\linewidth]{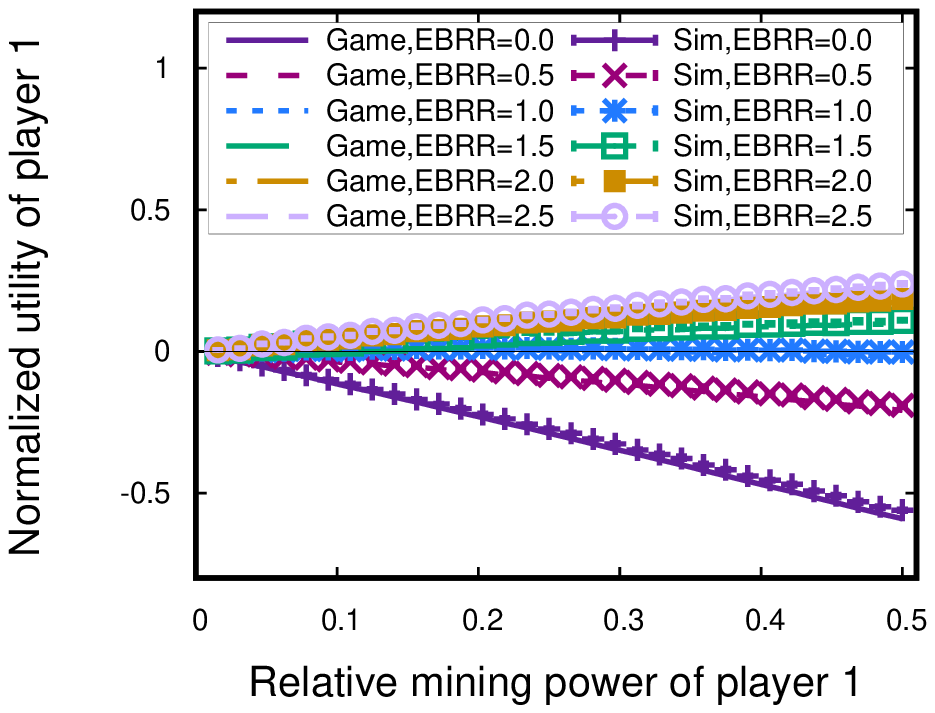}
                \caption{$ \SETTINGMEDOP $}
                \label{subfig:SimulationTheoryProfitComparisonB}
        \end{subfigure}%

        \begin{subfigure}[b]{\VERTICALSUBFIGSCALE\textwidth}
                \includegraphics[width=\linewidth]{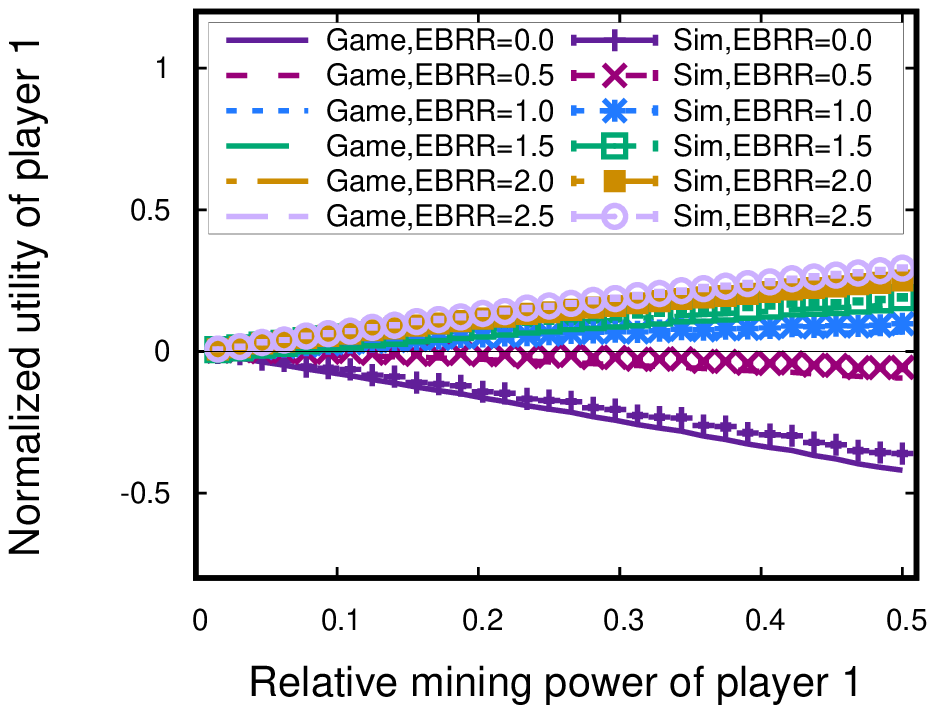}
                \caption{$ \SETTINGHIGHOP $}
                \label{subfig:SimulationTheoryProfitComparisonC}
        \end{subfigure}%
\negspace \negspace
        \caption{Comparison of the normalized utility of player~1~--- game analysis and simulation.}
        \label{fig:SimulationTheoryProfitComparison}
\end{figure}

We analyze a different scenario with arbitrary mining gaps. 
The game consists of two players that choose arbitrary start times for arbitrary portions of their rigs.
Each player partitions her controlled rigs into three sets, each with a different start time. 
We choose the start times arbitrarily, and their values are presented in Table~\ref{tab:splitting}.
Recall that $\rigRelativeStartTime{}{\rigIndex}$ is the start time of rig $\rigIndex$ normalized by the expected block time interval.
We use the game analysis and the simulator to obtain the normalized utility of player~1, and plot it as a function of her relative mining power.
The values of $\opexPerRigPerTime$ and $\capexPerRigPerTime$ are presented in Table~\ref{tab:opexCapexSettings}.
We repeat the analysis for different values of $\baseRewardRatio$.

Results are presented in Figure~\ref{fig:SimulationTheoryProfitComparison}.
As in the previous experiment, for the simulated results, we use the average of $\numberOfSimIterationsEachSetting$ different runs with different random seeds.
The error bars show the highest and lowest values.

This comparison demonstrates the effect of the $\baseRewardRatio$ value.
For the low values of~$\baseRewardRatio$, player~1 has negative utility.
As player~1 controls more rigs (i.e., has higher relative mining power), her per-mining-rig utility is decreasing with her total mining power.
Even though player~1 has higher probability to get rewarded as she controls more mining power, the increase in her expenses is more significant, resulting in lower utility.
For the higher values of $\baseRewardRatio$, the opposite occurs.
As player~1 controls more rigs, her per-rig-utility is increasing with her total mining power.
The increase in the probability to get rewarded surpasses the increase in expenses, resulting in higher utility.
This trend is maintained for all settings of opex and capex ratios.

Another interesting result shows the impact of the opex-capex ratio.
For any player~1 relative mining power and any $\baseRewardRatio$, the utility of player~1 where capex is dominant ($\SETTINGLOWOP$, Figure~\ref{subfig:SimulationTheoryProfitComparisonA}) is lower than when capex and opex and equal ($\SETTINGMEDOP$, Figure~\ref{subfig:SimulationTheoryProfitComparisonB}) and when opex is dominant ($\SETTINGHIGHOP$, Figure~\ref{subfig:SimulationTheoryProfitComparisonC}). 
Player~1's choice of start times that are greater than~0 is an optimization.
By doing so, she reduces her expected opex as her controlled rigs are expected to be active for less time. 
The more rigs she controls, the more impactful this effect is.
Hence, this suggests that at when opex is at play ($\SETTINGMEDOP$ ,$\SETTINGHIGHOP$), mining gap formation is beneficial. 

We conclude the simulations discussion with the following observation.
Recall the analysis is for the expected behavior and hence considers the expected pending transaction fees as part of the base reward. 
The simulations confirm the results predicted by the expected-case analysis, despite the fact the reward varies between individual blocks.

\snegspace
      \subsection{Case Studies} 
      \label{sec:caseStudies}
\snegspace

We present two insights regarding the optimal start time of players.
The first reviews, through an example, the effects of other players' start time strategies on the normalized utility of a player.
The second reviews optimal start times of players of different sizes.

\snegspace
\subsubsection{Case Study One~--- Effects of Other Players' Strategies}
\snegspace

\begin{figure}
		\centering
        \begin{subfigure}[b]{\VERTICALSUBFIGSCALE\textwidth}
                \includegraphics[width=\linewidth]{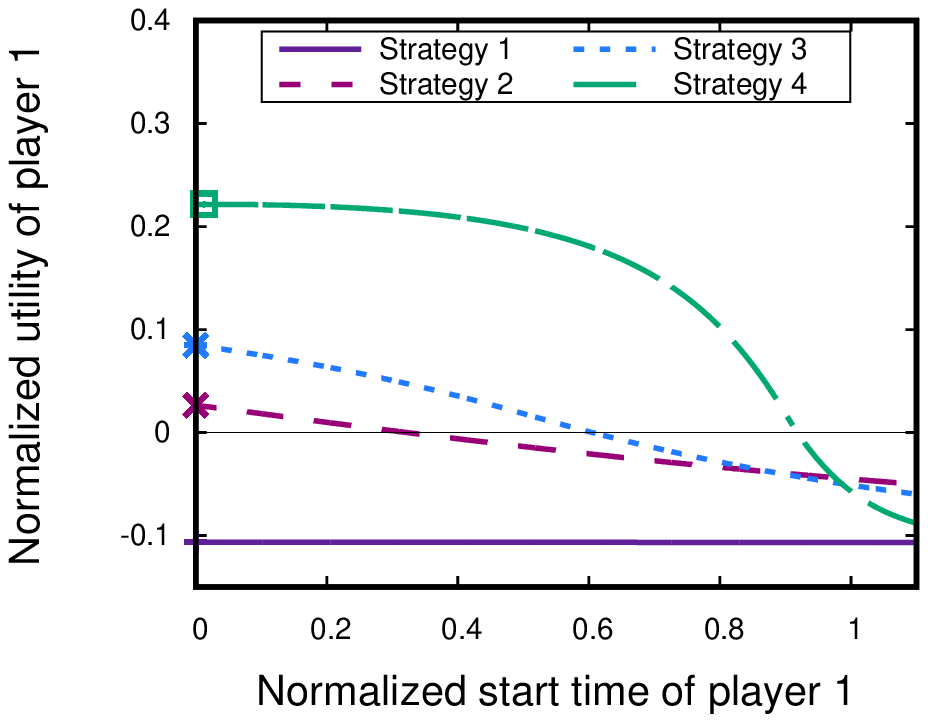}
                \caption{$ \SETTINGLOWOP $}
                \label{subfig:verifyProfitsA}
        \end{subfigure}%

        \begin{subfigure}[b]{\VERTICALSUBFIGSCALE\textwidth}
                \includegraphics[width=\linewidth]{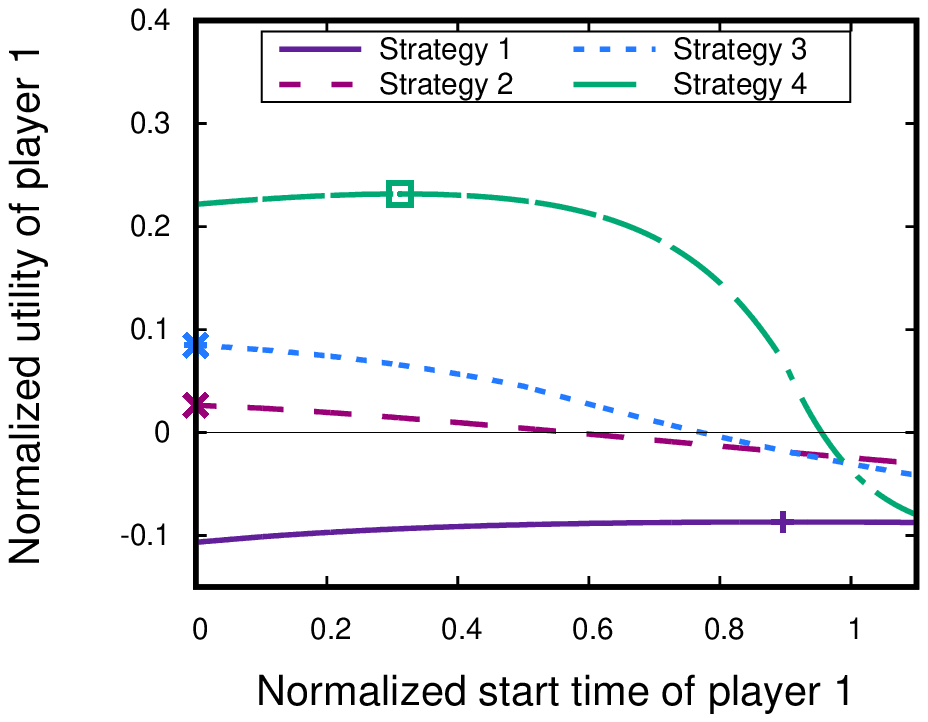}
                \caption{$ \SETTINGMEDOP $}
                \label{subfig:verifyProfitsB}
        \end{subfigure}%

        \begin{subfigure}[b]{\VERTICALSUBFIGSCALE\textwidth}
                \includegraphics[width=\linewidth]{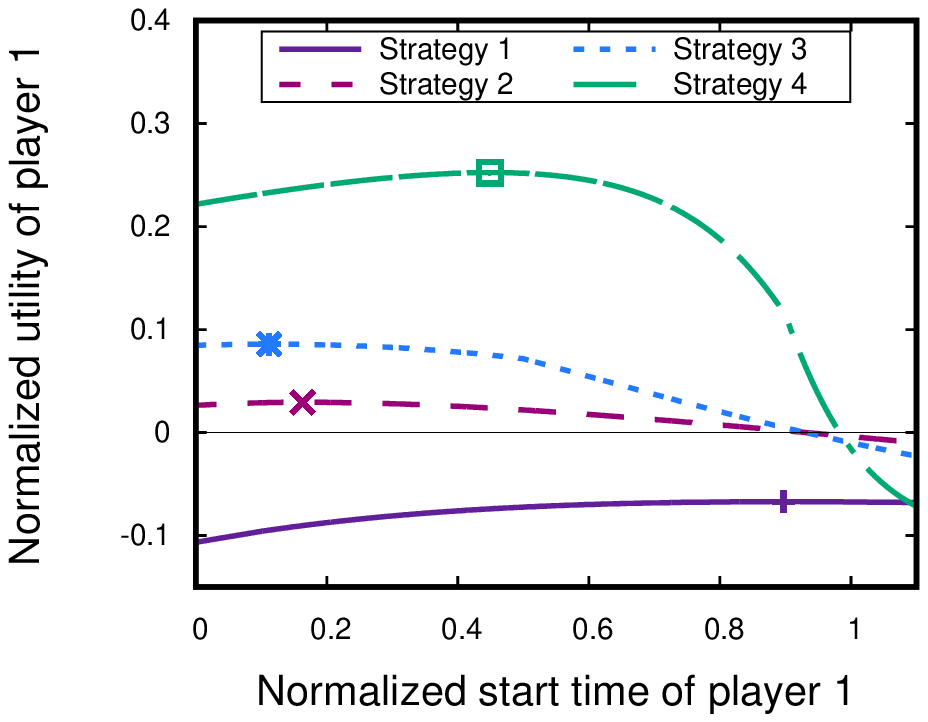}
                \caption{$ \SETTINGHIGHOP $}
                \label{subfig:verifyProfitsC}
        \end{subfigure}%
\negspace \negspace
        \caption{Normalized utility of player 1, for different strategies of other players.}
        \label{fig:utilityAsfunctionOfStartTime}
\end{figure}

\begin{table}
\centering
\begin{tabular}{ |c|c|c|c| } 
\hline
				  &  \multicolumn{3}{c|}{\textbf{Number of players}} \\
\hline
\textbf{Normalized start times}&  \textbf{0.1} & \textbf{0.5} & \textbf{0.9} \\
\hline
Strategy~1  					&  4    & 0    & 3 \\
\hline
Strategy~2  					&  7    & 0    & 0\\
\hline
Strategy~3  					&  0    & 7    & 0\\
\hline
Strategy~4  					&  0    & 0    & 7\\
\hline
\end{tabular}
\captionof{table}{Normalized start times of the other players.\negspace} 
\label{tab:othersStartTimeConfig} 
\end{table}

In our example we use a game with~8 players, controlling~16 rigs each, where each player selects a single mutual start time for all of her rigs.
We use $\baseRewardRatio=2$ for this example.
In Figure~\ref{fig:utilityAsfunctionOfStartTime} we present the normalized utility of player~1 as a function of her rigs start time for different start times of the other players.
The maximal value of each curve is marked.
Start times strategies of the other players are listed in Table~\ref{tab:othersStartTimeConfig}.
In strategy~1, four of the players choose normalized start time of~0.1 while the remaining three choose~0.9.
In strategies~2,3 and~4, all the other seven players choose normalized start times of~0.1,~0.5 and~0.9, respectively.

An increase in player's normalized utility is achieved by two means~--- increasing her chance of being rewarded and therefore increasing her expected reward, and by reducing her expenses. 
When a player chooses an early start time, she prefers to increase her chance for the reward, at the cost of increasing her expenses.
When a player chooses a late start time, she prefers to decrease her expenses, at the cost of lowering her chances to be rewarded.

Notice strategy~4, where all other seven players choose normalized start time of~0.9.
At the $ \SETTINGLOWOP $  setting, where $\opexPerRigPerTime=0$, player~1 can increase her chances of being rewarded without an increase in her expenses.
Hence, the optimal normalized start time as seen in Figure~\ref{subfig:verifyProfitsA} is zero.
At the $ \SETTINGMEDOP $  and  $ \SETTINGHIGHOP $ settings, where $\opexPerRigPerTime>0$, the conflict described above comes in play.
Choosing normalized start time of zero will cause unnecessary expenses, resulting in sub optimal normalized utility.
Choosing a relatively late normalized start time, such as~0.9, will also result in sub optimal normalized utility, as now player~1 has much lower chances to be rewarded and therefore much lower normalized utility.
The optimal normalized start time is therefore a time that balances the two conflicting interests.
From Figures~\ref{subfig:verifyProfitsB} and~\ref{subfig:verifyProfitsC} we can learn the optimal normalized start time in this case is in the range of $\left[0.2,0.5\right]$. 

At strategy~4, player~1 has relatively long period of time where she was the only player with active rigs.
This leads to relatively high chance for her to be rewarded, which she could forfeit to reduce her expenses. 
When the other players use strategy~2 for example, this privilege doesn't exist anymore, and player~1 shouldn't forfeit any chance she can get to win the reward. 
Hence, in all settings, her optimal normalized start time is~0.
Strategy~3 is in a sense the average case. 
The other players start at~0.5.
This start isn't too early yet not too late, and player~1 can optimize.
As expected, the optimal time is also dependent on the opex value.

Strategy~1 demonstrates the opposite case, where player~1 is better off waiting to decrease her expenses.
When $\opexPerRigPerTime>0$, player~1 minimizes her expenses by choosing fairly late start times.
When $\opexPerRigPerTime=0$, player~1 can't reduce expenses by choosing later start times, and therefore the optimal choice is normalized start time of zero.
\snegspace
\subsubsection{Case Study Two~--- Different-Size Players}
\snegspace

We now use the \optimizer{} to analyze a scenario with players of different sizes.
In this scenario we use the $\SETTINGHIGHOP$ setting with $\baseRewardRatio=2$.
We present the equilibria obtained by the \optimizer{} for some arbitrary sets of players.
Sets at examination and the resulting equilibria start times are presented in Table~\ref{tab:caseStudy}.

\begin{table}
\begin{tabular}{ |c|c|c|c|c| } 
\hline
				  &  \multicolumn{4}{c|}{\textbf{Relative Size, Normalized Start Time}} \\
\hline
\textbf{\#}			  &  \textbf{Player~1} & \textbf{Player~2} & \textbf{Player~3} & \textbf{Player~4} \\
\hline
1 & 0.125, \textbf{0.157} &  0.125, \textbf{0.157}  & 0.250, \textbf{0.261} & 0.5, \textbf{0.452} \\
\hline
2 & 0.250, \textbf{0.261} &  0.250, \textbf{0.261}  & 0.500, \textbf{0.452} & -, \textbf{-} \\
\hline
3 & 0.125, \textbf{0.131} &  0.375, \textbf{0.350}  & 0.500, \textbf{0.452} & -, \textbf{-} \\
\hline
4 & 0.125, \textbf{0.131} &  0.250, \textbf{0.261}  & 0.625, \textbf{0.452} & -, \textbf{-} \\
\hline
\end{tabular}
\captionof{table}{Case study of different size players.\negspace} 
\label{tab:caseStudy} 
\end{table}

We note that players with the same size choose same start times, such as player~1 and player~2 in scenario~1.
We also note that players with higher relative size choose higher start times.
Another result is that the bigger player in each scenario picks the same start time, even when the smaller other players choose different start times.

We now present an intuition for these results.
Consider a player of $1-\epsilon$ relative mining power for some infinitely small $\epsilon$.
This player is practically guaranteed to find the block and get the reward, whether she chooses early or late start times for her rigs.
Such player will then prefer to cut her expenses by choosing later start times, as her chances of winning are practically unaffected by such choice.
Now consider the opposite case, with a player of $\epsilon$ relative mining power.
This player has low chance to win the reward and she cannot afford to dwindle it any further.
Hence, such player will choose start time~0, to maintain what low chances of getting a reward she has. 
Intuitively, the higher relative power a player controls, the later the start time she prefers for her rigs.

We now move to analyze simpler settings, where all players are of equal sizes.

\snegspace
        \subsection{Equal-Size Miners Equilibria}
        \label{sec:equalPlayersEq}
\snegspace
We proceed to analyze equilibria where all miners are of equal sizes.
For a varying number of players, we divide the $\totalNumberOfRigs$ mining rigs among the players evenly, creating a set of equal-size players.
For the different settings presented in Table~\ref{tab:opexCapexSettings}, and different values of $\baseRewardRatio$, we use the \optimizer{} to find equilibria start times for the players.

\begin{figure}
		\centering
        \begin{subfigure}[b]{\VERTICALSUBFIGSCALE\textwidth}
                \includegraphics[width=\linewidth]{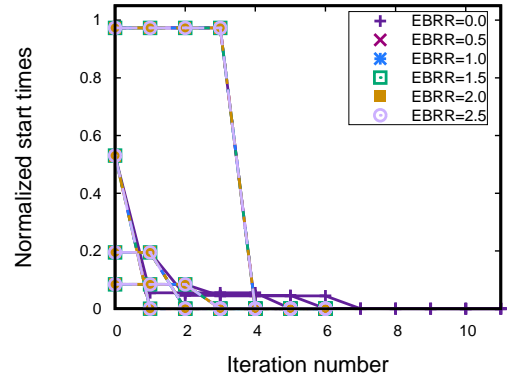}
                \caption{$ \SETTINGLOWOP $}
                \label{subfig:startTimeConvergenceA}
        \end{subfigure}%

        \begin{subfigure}[b]{\VERTICALSUBFIGSCALE\textwidth}
                \includegraphics[width=\linewidth]{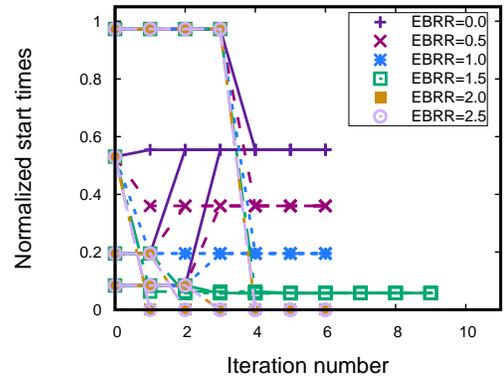}
                \caption{$ \SETTINGMEDOP $}
                \label{subfig:startTimeConvergenceB}
        \end{subfigure}%

        \begin{subfigure}[b]{\VERTICALSUBFIGSCALE\textwidth}
                \includegraphics[width=\linewidth]{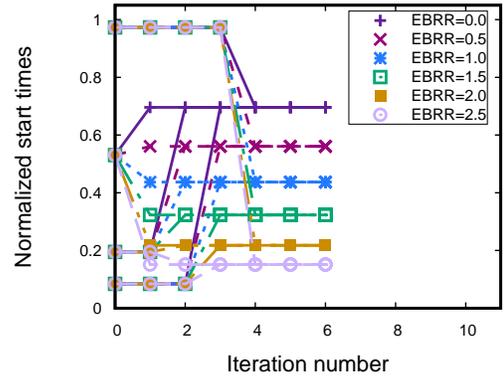}
                \caption{$ \SETTINGHIGHOP $}
                \label{subfig:startTimeConvergenceC}
        \end{subfigure}%
\negspace \negspace
        \caption{Convergence of normalized start times of equal-size players.}
        \label{fig:startTimeConvergence}
\end{figure}

We visualize, as an example, some of the equilibria search processes for a system comprised of~4 equal-size players, controlling~32 rigs each.
In Figure~\ref{fig:startTimeConvergence}, for the three different settings and different $\baseRewardRatio$ values, we plot at each iteration of the \optimizer{} the normalized start times of all of the~4 players.
We get the same qualitative results for any different numbers of players, and for different random initial start times.

We first notice that for all settings and for all values of $\baseRewardRatio$, each player eventually converges to the same start time.
We conclude symmetry holds.
We also notice that some settings require only one iteration before reaching the equilibrium start time, while other settings require a few iterations.
This strengthens the analysis result that the start times of other players affect the optimal strategy.
Another result is that different settings and values of $\baseRewardRatio$ lead to different optimal start times.
We discuss these result thoroughly in the following section.
\snegspace
            \subsubsection{Start Times at Equilibria}
\snegspace
\begin{figure}
		\centering
        \begin{subfigure}[b]{\VERTICALSUBFIGSCALE\textwidth}
                \includegraphics[width=\linewidth]{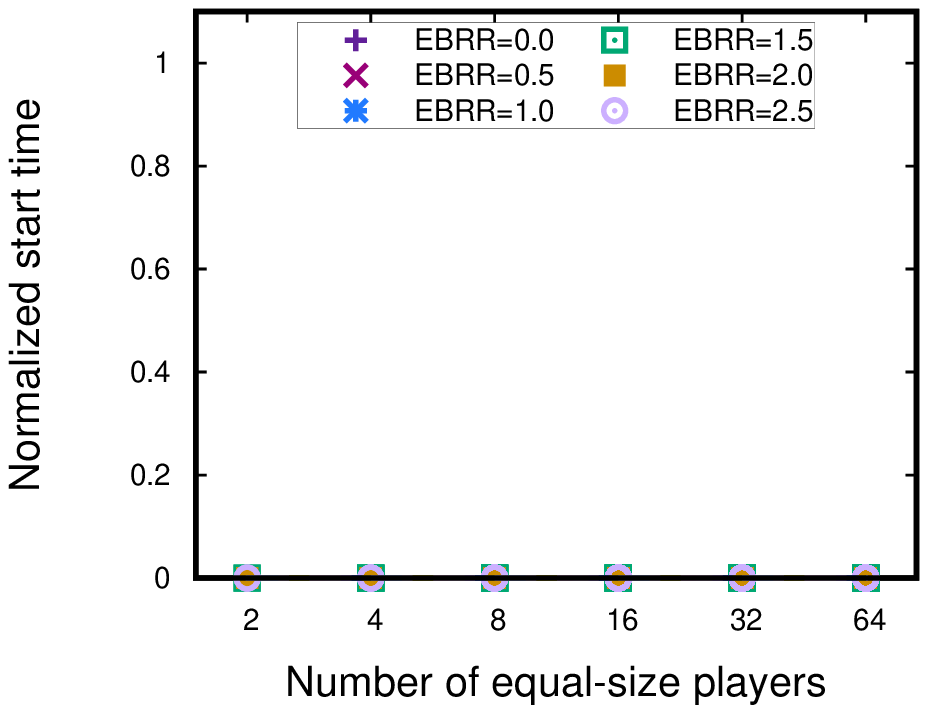}
                \caption{$ \SETTINGLOWOP $}
                \label{subfig:startTimePerSizeA}
        \end{subfigure}%

        \begin{subfigure}[b]{\VERTICALSUBFIGSCALE\textwidth}
                \includegraphics[width=\linewidth]{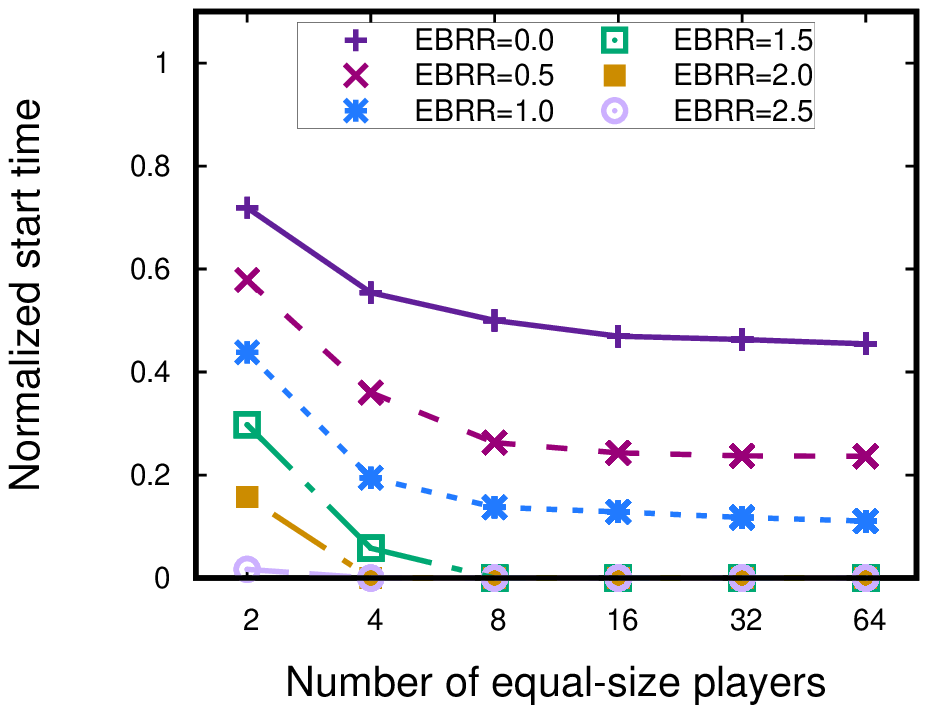}
                \caption{$ \SETTINGMEDOP $}
                \label{subfig:startTimePerSizeB}
        \end{subfigure}%

        \begin{subfigure}[b]{\VERTICALSUBFIGSCALE\textwidth}
                \includegraphics[width=\linewidth]{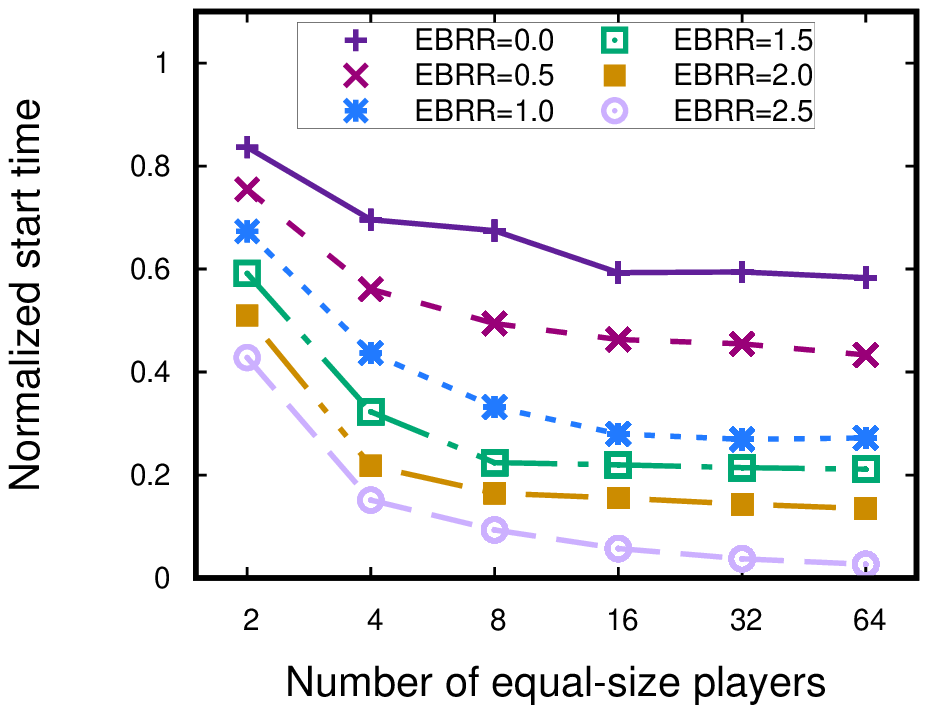}
                \caption{$ \SETTINGHIGHOP $}
                \label{subfig:startTimePerSizeC}
        \end{subfigure}%
\negspace \negspace
        \caption{Normalized start times of equal-size players.}
        \label{fig:startTimePerSize}
\end{figure}

In Figure~\ref{fig:startTimePerSize} we present the normalized start times at equilibrium of all miners as a function of the number of miners.
For the $ \SETTINGLOWOP $ setting, presented in Figure~\ref{subfig:startTimeConvergenceA}, the equilibrium is at start time zero for all values of $\baseRewardRatio$. 
This is expected, as $\opexPerRigPerTime=0$ and players do not suffer an increase in expenses by turning their rigs on earlier.
By setting their rigs' start times to zero, the players maximize their probability of getting rewarded, hence increasing their utility.
For the $ \SETTINGMEDOP $ and the $\SETTINGHIGHOP$ settings, presented in Figures~\ref{subfig:startTimeConvergenceB} and~\ref{subfig:startTimeConvergenceC} respectively, start times at equilibrium are zero only for the higher values of $\baseRewardRatio$.
When $\baseRewardRatio$ is low, the base reward is not substantial enough to incentivize players to choose start time zero, as they rather turn their rigs on at a later time and decrease their expected expenses.
Therefore the expenses prevented due to the optimization are more significant than the loss of potential reward.
When $\baseRewardRatio$ is high the base reward becomes more substantial and the opposite optimization takes place.
Players prefer start time zero, as the increase in probability to get the reward and therefore the expected reward are more significant than the increase in expenses.

Another interesting result is that players with higher relative power prefer later start times. 
An intuition for that was presented in Section~\ref{sec:caseStudies}.
For example, in a system with only~2 players, each player has a relative mining power of~0.5, these players choose the latest start time.
For systems with more players, say~16, each has relative mining power of~0.0625 and choose an earlier start time.
\snegspace
\subsubsection{Utility Increase from Optimization}
\snegspace
\begin{figure}
		\centering
        \begin{subfigure}[b]{\VERTICALSUBFIGSCALE\textwidth}
                \includegraphics[width=\linewidth]{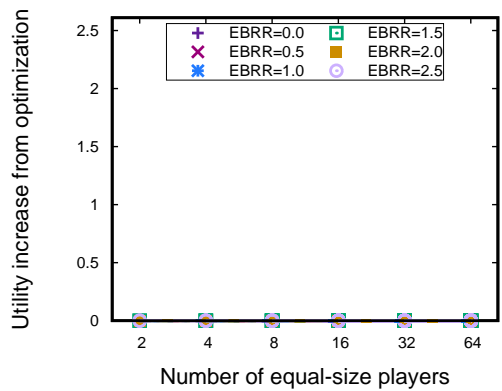}
                \caption{$ \SETTINGLOWOP $}
                \label{subfig:relativeProfitFromOptimizingA}
        \end{subfigure}%

        \begin{subfigure}[b]{\VERTICALSUBFIGSCALE\textwidth}
                \includegraphics[width=\linewidth]{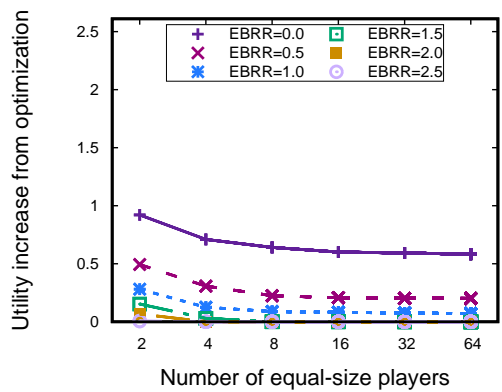}
                \caption{$ \SETTINGMEDOP $}
                \label{subfig:relativeProfitFromOptimizingB}
        \end{subfigure}%

        \begin{subfigure}[b]{\VERTICALSUBFIGSCALE\textwidth}
                \includegraphics[width=\linewidth]{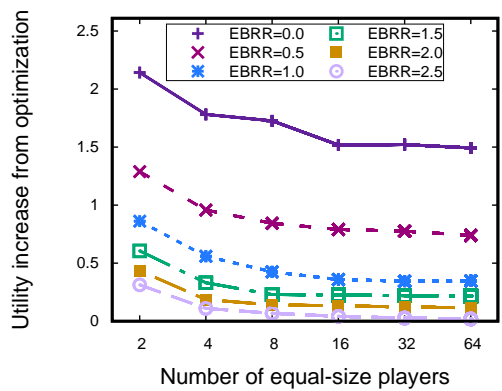}
                \caption{$ \SETTINGHIGHOP $}
                \label{subfig:relativeProfitFromOptimizingC}
        \end{subfigure}%
\negspace \negspace
        \caption{Utility increase from optimization.}
        \label{fig:relativeProfitFromOptimizing}
\end{figure}

In Figure~\ref{fig:relativeProfitFromOptimizing} we present the normalized utility increase of players from optimization. 
We measure the utility of players at the optimal and zero start times and subtract the latter from the former.

Recall that for the $ \SETTINGLOWOP $ setting, the equilibrium start time is zero, hence there is no increase in utility. 
This result is presented in Figure~\ref{subfig:relativeProfitFromOptimizingA},
For the $ \SETTINGMEDOP $ and the $\SETTINGHIGHOP$ settings, equilibrium start time is zero only for the higher values of $\baseRewardRatio$.
The results of such optimization are presented in Figures~\ref{subfig:relativeProfitFromOptimizingB} and~\ref{subfig:relativeProfitFromOptimizingC}.
When players optimize, they gain a substantial increase in their utility.
Notice that three factors contribute to an increase in utility~--- low $\baseRewardRatio$, high $\opexPerRigPerTime$ and a small number of players.
All these three make optimization more profitable, by reducing the expected reward from finding a block and increasing the potential gain of saving expenses.

\snegspace
\subsubsection{Mining Power Utilization}
\snegspace
\begin{figure}
		\centering
        \begin{subfigure}[b]{\VERTICALSUBFIGSCALE\textwidth}
                \includegraphics[width=\linewidth]{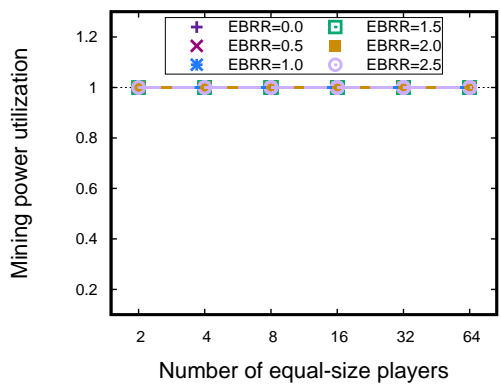}
                \caption{$ \SETTINGLOWOP $}
                \label{subfig:effectiveMiningPowerPerSizeA}
        \end{subfigure}%

        \begin{subfigure}[b]{\VERTICALSUBFIGSCALE\textwidth}
                \includegraphics[width=\linewidth]{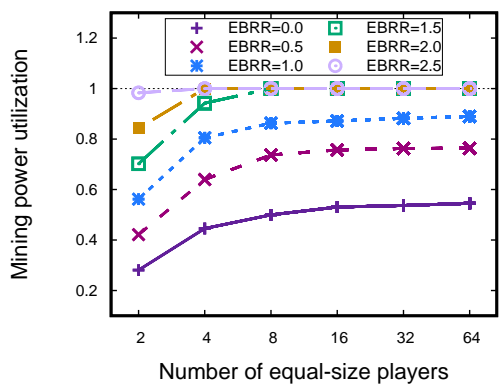}
                \caption{$ \SETTINGMEDOP $}
                \label{subfig:effectiveMiningPowerPerSizeB}
        \end{subfigure}%

        \begin{subfigure}[b]{\VERTICALSUBFIGSCALE\textwidth}
                \includegraphics[width=\linewidth]{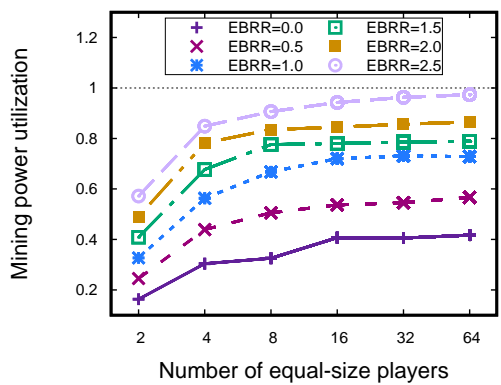}
                \caption{$ \SETTINGHIGHOP $}
                \label{subfig:effectiveMiningPowerPerSizeC}
        \end{subfigure}%
\negspace \negspace
        \caption{Mining power utilization.}
        \label{fig:effectiveMiningPowerPerSize}
\end{figure}

Equilibria with positive gap sizes negatively affect system security by reducing the amount of resources protecting the system. 
Recall that in proof of work systems, the security of the system relies on the honest miners' mining power.
When less mining power takes part, the system becomes less resilient to attacks, as now attackers require less resources. 
The mining power utilization~\cite{eyal2016ng} is the ratio of mining power that effectively secures the blockchain out of all mining power in the hands of well-behaved miners. 
If the mining power utilization is smaller than one, then an attacker can perform a~$51\%$ attack with less than~$51\%$ of the mining power, and selfish mining becomes easier to achieve. 
Figure~\ref{fig:effectiveMiningPowerPerSize} show the mining power utilization in various scenarios.
In Figure~\ref{subfig:effectiveMiningPowerPerSizeA}, when the $ \SETTINGLOWOP $ setting applies, all players choose start time zero, and the mining power utilization is not affected.
In Figures~\ref{subfig:effectiveMiningPowerPerSizeB} and~\ref{subfig:effectiveMiningPowerPerSizeC}, when the $ \SETTINGMEDOP $ and the $\SETTINGHIGHOP$ settings apply, players use mining gaps, leading to decrease in the mining power utilization.
Note that at the most extreme scenario of two players, high opex and low base reward, the mining power utilization drops to about~$10\%$.
\snegspace
\subsubsection{$\baseRewardRatio$ for a Limited Mining Gap}
\label{sec:baseRewardForNoGapEqualSize}
\snegspace

\begin{figure}[!t]
  \centering
  \includegraphics[width=\FIGSCALE\textwidth]{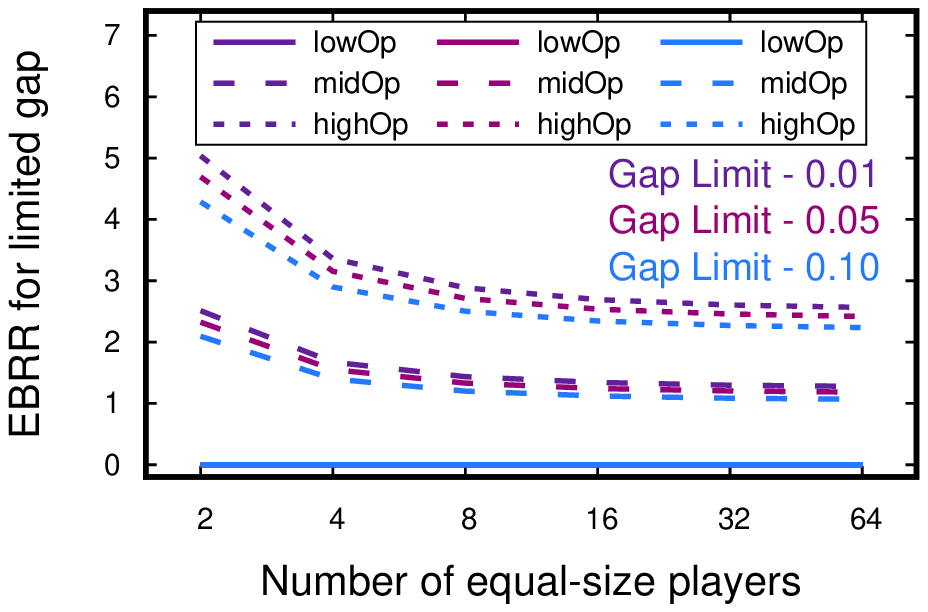} 
  \negspace \snegspace
  \caption{Minimal $\baseRewardRatio$ for a limited gap.}
 \label{fig:minBrrForNoGapPerSize}
\end{figure}

We have seen the implications on security from mining gaps, therefore we explore the question of how to avoid such gaps.
We find the minimal $\baseRewardRatio$ to limit the size of a mining gap.
Assume we want to limit the start time of players at equilibrium to be a factor of $x$ from the $\timeBetweenBlocks$.
Therefore, we look for the minimal $\baseRewardRatio$ value such that the start time at equilibrium will be less than $x \cdot \timeBetweenBlocks$.

We use binary search over a wide range of $\baseRewardRatio$ values and mark the lowest $\baseRewardRatio$ that bounds the gap by $x \cdot \timeBetweenBlocks$, for three different factors $\forall x \in \left\{0.01,0.05.0.1 \right\}$.
We repeat this experiment for various number of players as well as different opex and capex values.
We present the results in Figure~\ref{fig:minBrrForNoGapPerSize}.

As expected, for the $\SETTINGLOWOP$ setting, the start time at equilibrium is zero as mining is free, and even $\baseRewardRatio = 0$ suffices. 
As optimization becomes more profitable due to the aforementioned reasons, higher $\baseRewardRatio$ values are required to limit start times at the equilibrium.
When the $\baseRewardRatio$ is higher, the total reward from finding the block is higher.
Hence, it incentivizes miners to prefer increasing their chances of winning the reward over decreasing their expenses, which ultimately leads to a limited mining gap.

Note that even as the number of players grows, the curves converge to a fixed value of $\baseRewardRatio$.
We deduce that even in a system with many small miners, a gap still forms in the presence of opex.

\snegspace

\subsection{Case Study: Bitcoin}
\label{sec:caseStudyBitcoin}
\snegspace

We now make an educated estimation to when Bitcoin becomes prone to the undesired effects of mining gaps.
There are many operational cryptocurrency systems, all vary in minting, fees, market cap, and expenses.
Given such parameters for any cryptocurrency, a similar estimation can be performed using our model.
We present a case study of Bitcoin.

We consider the popular mining rig \emph{Antminer~s9}~\cite{antminers9Spec} with an estimated life expectancy of one year.
Its required power is about~1.3kW and average cost about~\$1000.
Electricity cost is about~\$0.1/kWh~\cite{wiki2018electricityPricing}.
In one year the electricity expenses of one miner sum up to~\$876, which means the system falls in the area of $\SETTINGMEDOP$.

In Bitcoin today there are~7 mining pools~\cite{blockchain2018bitcoinPools} controlling about~85\% of the mining power, while the rest is divided among many smaller mining pools.
Although they vary in size, we approximate that situation by assuming~8 equal-size miners.

Using results of Section~\ref{sec:baseRewardForNoGapEqualSize}, we deduce that $\baseRewardRatio \approx 1$ is required to maintain a small gap. 
Currently, the rewards from minting and fees are~\btc{}12.5 and about~\btc{}1, respectively.
Therefore currently $\baseRewardRatio \approx 12.5$, so gaps are not profitable.
However, in about ten years the minting reward drop will drop to about~1, which means $\baseRewardRatio \approx 1$ and the system will be in a state where gaps are profitable.

The ten-year estimate is an optimistic one, as it assumes the reward from fees does not increase. 
Different mining hardware, change in electricity costs, and changes in the currency market, all might lead to different results. 
We emphasize that our estimation does not consider incentives external to our model resulting in seemingly altruistic behavior~\cite{badertscher2018but}.

%
%

\section{Conclusion}
\label{sec:conclusion}

We defined and analyzed the gap game exploring how mining gaps form as a function of subsidy and fees, capex and opex. 
We showed that once fees become significant gaps form, though not uniformly as previous believed, and their effect on blockchain security is significant, decreasing mining utilization by up to $90\%$ in extreme scenario, and leading to centralization incentives. 

This means that base rewards are critical for system security, and should be achieved either by subsidy, fee backlogs, or alternative fee schemes~\cite{lavi2017redesigning,pass2017fruitchains}. 
We show that $\baseRewardRatio \approx 6$ is sufficient to avoid mining gaps in presented scenarios; we expect Bitcoin to drop below this threshold within a decade. 

Establishing that gaps occur is an early and important step in the security analysis of cryptocurrency systems.
This work is a step in that direction, demonstrating that gap analysis is critical for a more complete security analysis of blockchains. 
Such analysis can be used to inform the design of future and current cryptocurrencies. 

\subsection*{Acknowledgements}

This research was supported by the Israel Science Foundation (grant No. 1641/18), the Technion Hiroshi Fujiwara cyber-security research center, and the Israel cyber bureau.

\bibliographystyle{ACM-Reference-Format}
\bibliography{btc}

\appendix

\section{Valid PDF Proof}
\label{app:validPdf}

Denote $\startTimesIntervalSet{}$ the list of all start time intervals.
Note that for any interval $(\rigStartTime{}{\otherRigIndex},\rigStartTime{}{\otherRigIndex+1}) \in \startTimesIntervalSetName{}$, no rigs are turned on, meaning $\allOperatingRigsAllPlayers{}$ does not change.

Our goal is to show that $\int_{-\infty}^{\infty} \pdfRigMin d \currentTime = 1$.
We begin by taking notice that at time ${\rigStartTimeVectorWithIndex{\otherRigIndex+1}}$ rig $\otherRigIndex + 1$ becomes active, 
resulting in
$\allOperatingRigsAllPlayersAtSomeStartTimeWithRigIndex{
\rigStartTimeVectorWithIndex{\otherRigIndex+ 1}} 
\setminus 
\allOperatingRigsAllPlayersAtSomeStartTimeWithRigIndex{
\rigStartTimeVectorWithIndex{\otherRigIndex}} 
= \left\{\rigStartTime{}{\otherRigIndex + 1}\right\}$.
We get that for any~$\otherRigIndex \in \allRigsIndices$ :
\begin{equation}
\label{eq:pdfValidationLemma}
\begin{split}
{}&		\sumofTimeSinceItemsInSetAtTimeWithVal{
 		{\rigStartTimeVectorWithIndex
 		{\otherRigIndex+1}}}{\rigStartTimeVectorWithIndex{\otherRigIndex+1}} =\\
{}&	 		 \sumofTimeSinceItemsInSetAtTimeWithVal{
 		{\rigStartTimeVectorWithIndex
 		{\otherRigIndex}}}{\rigStartTimeVectorWithIndex{\otherRigIndex+1}} + 
 		{\rigStartTimeVectorWithIndex{\otherRigIndex + 1}} -
 		\rigStartTime{}{\otherRigIndex + 1} =   \\
{}&	 		 \sumofTimeSinceItemsInSetAtTimeWithVal{
 		{\rigStartTimeVectorWithIndex
 		{\otherRigIndex}}}{\rigStartTimeVectorWithIndex{\otherRigIndex+1}} + 
 		\rigStartTime{}{\otherRigIndex + 1} -
 		\rigStartTime{}{\otherRigIndex + 1} =   \\ 		
{}&	 		\sumofTimeSinceItemsInSetAtTimeWithVal{
 		{\rigStartTimeVectorWithIndex
 		{\otherRigIndex}}}{\rigStartTimeVectorWithIndex{\otherRigIndex+1}}\,\,\,\,\,\,\,\,\,\,\,.		
\end{split}
\end{equation}

We are now ready to present the full verification process, which is detailed in Equation~\ref{eq:pdfRigMinVerify}.

\begin{figure*}[!t] 
 \centering

 \begin{subfigure}[b]{\textwidth}
\label{fig:eq:pdfRigMinVerifyPartOne}
\begin{equation}
\label{eq:pdfRigMinVerify}
\begin{split}
{}&\int_{-\infty}^{\infty} \pdfRigMin d \currentTime = \\
{}& \sum_{(\rigStartTime{}{\otherRigIndex},\rigStartTime{}{\otherRigIndex+1}) \in \startTimesIntervalSetName{}}
 {\left [ 
 		\int_{\rigStartTimeVectorWithIndex{\otherRigIndex}}
 		^{\rigStartTimeVectorWithIndex{\otherRigIndex + 1}}  
 		{\pdfRigMin d \currentTime} 
 \right ]}= \\
 {}& \sum_{(\rigStartTime{}{\otherRigIndex},\rigStartTime{}{\otherRigIndex+1}) \in \startTimesIntervalSetName{}}
 {\left [
 		\int_{\rigStartTimeVectorWithIndex{\otherRigIndex}}
 		^{\rigStartTimeVectorWithIndex{\otherRigIndex + 1}}  
 		{ \difficultyPerRig \cdot \sizeOfSet{\allOperatingRigsAllPlayers{}} \cdot
   		\exp\left({-\difficultyPerRig \cdot \sumofTimeSinceItemsInSet{}}\right)} 
 \right ]} = \\
{}& \sum_{(\rigStartTime{}{\otherRigIndex},\rigStartTime{}{\otherRigIndex+1}) \in \startTimesIntervalSetName{}}
 {\left [ 
 		\int_{\rigStartTimeVectorWithIndex{\otherRigIndex}}
 		^{\rigStartTimeVectorWithIndex{\otherRigIndex + 1}}  
 		{    \difficultyPerRig \cdot
 		\sizeOfSet{\allOperatingRigsAtSomeStartTime{{\rigStartTimeVectorWithIndex{\otherRigIndex}}}} 
 		\cdot \exp\left({-\difficultyPerRig \cdot
 		\sumofTimeSinceItemsInSetAtTimeWithCurrentTime{{\rigStartTimeVectorWithIndex{\otherRigIndex}}}}\right)}
 \right ]} = \\ 
{}& \sum_{(\rigStartTime{}{\otherRigIndex},\rigStartTime{}{\otherRigIndex+1}) \in \startTimesIntervalSetName{}}
 \left[
 		\left (  
 			{- \exp\left({-\difficultyPerRig \cdot
 			\sumofTimeSinceItemsInSetAtTimeWithCurrentTime{{\rigStartTimeVectorWithIndex{\otherRigIndex}}}}\right)}
 			\Biggr|_{\rigStartTimeVectorWithIndex{\otherRigIndex}}
 			^{\rigStartTimeVectorWithIndex{\otherRigIndex+1}}
 		\right)
 \right ] =  \\
{}& \sum_{(\rigStartTime{}{\otherRigIndex},\rigStartTime{}{\otherRigIndex+1}) \in \startTimesIntervalSetName{}}
\left [  
 		{ \exp\left({-\difficultyPerRig \cdot
 		\sumofTimeSinceItemsInSetAtTimeWithVal{
 		{\rigStartTimeVectorWithIndex
 		{\otherRigIndex}}}{\rigStartTimeVectorWithIndex{\otherRigIndex}}}\right)} 
 		{- \exp\left({-\difficultyPerRig \cdot
 		\sumofTimeSinceItemsInSetAtTimeWithVal{
 		{\rigStartTimeVectorWithIndex
 		{\otherRigIndex}}}{\rigStartTimeVectorWithIndex{\otherRigIndex+1}}}\right)} 
 \right ] = \\
\shortintertext{Using Equation~\ref{eq:pdfValidationLemma} we get the last expression is a telescopic sum. Substituting in the relevant expressions yields} 
{}& \exp\left({-\difficultyPerRig \cdot
 		\sumofTimeSinceItemsInSetAtTimeWithVal{
 		{\rigStartTimeVectorWithIndex
 		{1}}}{\rigStartTimeVectorWithIndex{1}}}\right) - 
 		\exp\left({-\difficultyPerRig \cdot
 		\sumofTimeSinceItemsInSetAtTimeWithVal{
 		{\rigStartTimeVectorWithIndex
 		{\totalNumberOfRigs }}}{\infty}}\right) = \\
{}& \exp
 		\left(
 		0
 		\right)  
 		-\exp
 		\left(
		- \infty
 		\right) =  1 - 0 = 1 \shortintertext{as required.} 
\end{split}
\end{equation}
\caption{Verifying the PDF.}
\end{subfigure}

 \begin{subfigure}[b]{\textwidth}
\begin{equation}
\label{eq:blockTimeExpressionForDiff}
\begin{split}
{}&\timeBetweenBlocks = \meanValOfBlockFindingRandomEvent =  
	 \int_{0}^{\infty} \probabilityTimeLessThanBlock d \currentTime = \\
{}& \int_{0}^{\rigStartTimeVectorWithIndex{1}} 
			{1 d \currentTime} + 
		 \sum_{(\rigStartTime{}{\otherRigIndex},\rigStartTime{}{\otherRigIndex+1}) \in \startTimesIntervalSetName{}}
			 {\left [ 
			 		\int_{\rigStartTimeVectorWithIndex{\otherRigIndex}}
			 		^{\rigStartTimeVectorWithIndex{\otherRigIndex + 1}}  
			 		{\exp\left({-\difficultyPerRig \sumofTimeSinceItemsInSet{} }\right) d \currentTime} 
			 \right ]}= \\
{}& \rigStartTimeVectorWithIndex{1} + 
		 \sum_{(\rigStartTime{}{\otherRigIndex},\rigStartTime{}{\otherRigIndex+1}) \in \startTimesIntervalSetName{}}
			 {\left [ 
			 		\int_{\rigStartTimeVectorWithIndex{\otherRigIndex}}
			 		^{\rigStartTimeVectorWithIndex{\otherRigIndex + 1}}  
			 		{\exp\left({-\difficultyPerRig \sumofTimeSinceItemsInSet{} }\right) d \currentTime} 
			 \right ]}= \\
{}& \rigStartTimeVectorWithIndex{1} + 
		\sum_{(\rigStartTime{}{\otherRigIndex},\rigStartTime{}{\otherRigIndex+1}) \in \startTimesIntervalSetName{}}
		 {\left [ 
		 		\int_{\rigStartTimeVectorWithIndex{\otherRigIndex}}
		 		^{\rigStartTimeVectorWithIndex{\otherRigIndex + 1}}  
		 		{    \exp\left({-\difficultyPerRig \cdot
		 		\sumofTimeSinceItemsInSetAtTimeWithCurrentTime{{\rigStartTimeVectorWithIndex{\otherRigIndex}}}}\right)}
		 \right ]} = \\ 
{}& \rigStartTimeVectorWithIndex{1} + 
			\sum_{(\rigStartTime{}{\otherRigIndex},\rigStartTime{}{\otherRigIndex+1}) \in \startTimesIntervalSetName{}} 
			\left[
			 		\left (  
			 			{- \dfrac{1}
			 			{ \difficultyPerRig \cdot
		 				\sizeOfSet{\allOperatingRigsAtSomeStartTime{{\rigStartTimeVectorWithIndex{\otherRigIndex}}}}
		 				}
			 			 \exp\left({-\difficultyPerRig \cdot
			 			\sumofTimeSinceItemsInSetAtTimeWithCurrentTime{{\rigStartTimeVectorWithIndex{\otherRigIndex}}}}\right)}
			 			\Biggr|_{\rigStartTimeVectorWithIndex{\otherRigIndex}}
			 			^{\rigStartTimeVectorWithIndex{\otherRigIndex+1}}
			 		\right)
			 \right ] = \\
{}& \rigStartTimeVectorWithIndex{1} + 
		\sum_{(\rigStartTime{}{\otherRigIndex},\rigStartTime{}{\otherRigIndex+1}) \in \startTimesIntervalSetName{}}
	 			{ \dfrac{
		 		{ \exp\left({-\difficultyPerRig \cdot
		 		\sumofTimeSinceItemsInSetAtTimeWithVal{
		 		{\rigStartTimeVectorWithIndex
		 		{\otherRigIndex}}}{\rigStartTimeVectorWithIndex{\otherRigIndex}}}\right)}
		 		{- \exp\left({-\difficultyPerRig \cdot
		 		\sumofTimeSinceItemsInSetAtTimeWithVal{
		 		{\rigStartTimeVectorWithIndex
		 		{\otherRigIndex}}}{\rigStartTimeVectorWithIndex{\otherRigIndex+1}}}\right)} }
	 			{ \difficultyPerRig \cdot
 				\sizeOfSet{\allOperatingRigsAtSomeStartTime{{\rigStartTimeVectorWithIndex{\otherRigIndex}}}}
 				}}\,.
\end{split}
\end{equation}
\caption{Difficulty parameter $\difficultyPerRig$ at equilbria constraint. }
\end{subfigure}

\end{figure*}

\section{Difficulty Parameter Value at Equilibrium}
\label{app:calcDiff}

In this section we show how to find the value of $\difficultyPerRig$ at equilibrium.
The system's protocol dictates that the mean block creation time interval is $\timeBetweenBlocks$.
This is done by setting the value of $\difficultyPerRig$.
At equilibrium, the mean block creation time interval is the expected time to find a block $\meanValOfBlockFindingRandomEvent$.
Hence at equilibrium $\meanValOfBlockFindingRandomEvent = \timeBetweenBlocks$.
We use this equality to express a constraint on the system at equilibrium.
A known result in probability theory applies here~--- since $\blockFindingRandomEvent$ is a non-negative random variable, its expected value $\meanValOfBlockFindingRandomEvent = \int_{0}^{\infty} \probabilityTimeLessThanBlock d \currentTime $.
This results in
\begin{equation*}
\begin{split}
 \timeBetweenBlocks &= \\
 									  &= \meanValOfBlockFindingRandomEvent  \\
 									  &= \int_{-\infty}^{\infty} \left(\currentTime \pdfRigMin\right) d \currentTime\\
 									  &= \int_{0}^{\infty} \left(1 - \cdfRigMin \right) d \currentTime\\		
 									  &= \int_{0}^{\infty} \probabilityTimeLessThanBlock d \currentTime \,.
\end{split}
\end{equation*}

Based on Equation~\ref{eq:cdfBlock}, we know that~$ \probabilityTimeLessThanBlock $ is equal to $  \exp\left({-\difficultyPerRig \sumofTimeSinceItemsInSet{} }\right)$.
We again use $\startTimesIntervalSetName{}$ notion that was defined in Appendix~\ref{app:validPdf}
Note that for any interval $(\rigStartTime{}{\otherRigIndex},\rigStartTime{}{\otherRigIndex+1}) \in \startTimesIntervalSetName{}$, no rigs are turned on, meaning $\allOperatingRigsAllPlayers{}$ does not change.
In Equation~\ref{eq:blockTimeExpressionForDiff} we derive an expression for $\meanValOfBlockFindingRandomEvent$ as a function of $\difficultyPerRig$ and $\rigStartTimeVector$.
Note that this is an implicit function with respect to $\difficultyPerRig$.

\end{document}